\PassOptionsToPackage{sort&compress}{natbib}
\documentclass[preprint,12pt]{elsarticle}

\usepackage{amsmath,amsfonts,amssymb}  
\usepackage{mathptmx}                  
\usepackage{graphicx}                  
\usepackage{array}                     
\usepackage{booktabs}                  
\usepackage{url}                       
\usepackage{textcomp}                  
\usepackage{stfloats}                  
\usepackage{verbatim}                  
\usepackage{xspace}                    
\usepackage{xcolor}                  
\usepackage[most]{tcolorbox}
\tcbuselibrary{skins}
\usepackage{verbatim}
\usepackage{tabularx}
\usepackage{multirow}
\usepackage{siunitx} 
\usepackage{listings}
\usepackage[caption=false,font=normalsize,labelfont=sf,textfont=sf]{subfig}
\usepackage{enumitem}

\usepackage[normalem]{ulem}            
\usepackage{fontawesome}               
\usepackage{cuted}                     

\usepackage{algorithm}
\usepackage{algorithmic}

\usepackage{lineno}

\definecolor{ent}{HTML}{606266}
\definecolor{col}{HTML}{DAA520}

\newcommand{\xbRevvv}[1]{{\color{black} #1}}    
\newcommand{\rev}[1]{{\color{black} #1}}    
\newcommand{\virev}[1]{{\color{black} #1}}    
\newcommand{\mrev}[1]{{\color{black} #1}}    
\newcommand{\xb}[1]{{\color{black} #1}}
\newcommand{\xbNote}[1]{{\color{black} #1}}      

\newcommand{\lsy}[1]{{\color{black} #1}}
\newcommand{\lsyRev}[1]{{\color{black} #1}}    

\newcommand{\ie}{i.e.}
\newcommand{\eg}{e.g.}

\newcommand{\imp}[1]{\textbf{\textit{#1}}}
\newcommand{\example}[1]{\emph{``#1''}}

\newcommand{\systemname}{{QRec-NLI}}
\newcommand{\name}{{\textit{QRec-NLI}}}

\newcommand{\ui}{{\textit{User Interface}}}
\newcommand{\rec}{{\textit{Recommendation Engine}}}
\newcommand{\nlsql}{{\textit{Query Analyzer}}}
\newcommand{\vg}{{\textit{Visualization Generator}}}
\newcommand{\qlog}{{\textit{Query Log Repository}}}
\newcommand{\dbs}{{\textit{Database Selector}}}
\newcommand{\ti}{{\textit{Table Info}}}
\newcommand{\hp}{{\textit{History Panel}}}
\newcommand{\qp}{{\textit{Query Panel}}}
\newcommand{\rp}{{\textit{Result Panel}}}

\newenvironment{compactitem}
  {\begin{itemize}[nosep,leftmargin=*]}
  {\end{itemize}}

\newtcolorbox{casebox}{
  breakable,         
  colback=white,     
  colframe=black,    
  boxrule=1pt,     
  arc=2mm,           
  before skip=\medskipamount, 
  after skip=\medskipamount,
}

\usepackage[pagebackref,bookmarks]{hyperref}

\newcommand{\code}[1]{\texttt{#1}}
\newtcolorbox{exampleblock}[2][]{%
  enhanced,
  breakable,
  colback=white,
  colframe=black!15,
  coltitle=black,
  fonttitle=\normalsize,
  attach boxed title to top left={yshift=-1pt, xshift=1.5mm},
  boxed title style={colback=white, boxrule=0pt, frame empty},
  borderline west={2pt}{0pt}{blue!55},
  sharp corners,
  left=1.5mm,right=1.5mm,top=0.4ex,bottom=0.4ex,
  boxsep=0.4ex,
  before skip=2pt, after skip=2pt,
  noparskip,
  before upper=\vspace{-1pt},
  after upper=\vspace{-1pt},
  #1
}

\journal{Visual Informatics}

\begin{document}

\begin{frontmatter}



\title{Beyond Interestingness: Semantic and Context-Aware Natural
Language Query Recommendations for \\ Visual Data Analysis} 

\author[bcai,hkust]{Xingbo Wang} 
\tnotetext[hkust]{Work done during Xingbo Wang and Furui Cheng conducted PhD studies at HKUST.}
\author[dut]{Siyuan Li} 
\author[hkust]{Furui Cheng} 
\author[ntu]{Yong Wang}
\author[hw]{Jiang Long}
\author[hw]{Hong Lu}
\author[hkust]{Huamin Qu} 
\author[nju]{Ke Xu}

\affiliation[bcai]{organization={Present Address: Bosch Research North America \& Bosch Center for Artificial Intelligence (BCAI)},
            city={Sunnyvale},
            state={CA},
            country={USA}}

\affiliation[dut]{organization={Dalian University of Technology},
            city={Dalian},
            state={Liaoning},
            country={China}}


\affiliation[hkust]{organization={Department of Computer Science and Engineering, The Hong Kong University of Science and Technology (HKUST)},
            city={Hong Kong},
            state={Hong Kong},
            country={China}}

\affiliation[hw]{organization={Technical Innovation Department, Huawei Technologies Co. Ltd.},
            city={Hangzhou},
            state={Zhejiang},
            country={China}}

\affiliation[ntu]{organization={College of Computing and Data Science, Nanyang Technological University},
            city={Singapore},
            country={Singapore}}

\affiliation[nju]{organization={Nanjing University},
            city={Nanjing},
            state={Jiangsu},
            country={China}}

\begin{abstract}
\rev{Recent advances in large language models (LLMs) have made natural language interfaces (NLIs) widely accessible for data exploration, 
\virev{yet analysts who have a broad analytical objective still face the challenge of decomposing it into effective step-by-step queries, especially over unfamiliar, multi-table relational databases.}
\virev{Rather than generating high-level analytical agendas,}
we investigate how to augment an NLI with semantic- and context-aware next-step query recommendations that act as analytical scaffolding for relational database exploration. 
Our approach goes beyond interestingness-only methods by jointly integrating semantic relevance, data interestingness, and context coherence to guide exploration toward coherent, topic-focused analyses and potentially insightful subsets.
We evaluate \systemname{} with NL2SQL benchmarking, LLM-enhanced description validation, agentic comparisons against interestingness-only and LLM-based prompting baselines, and a 12-participant user study.} \mrev{In the agentic comparison, \systemname{} yields more topically relevant and locally coherent query sequences than both baselines. In the user study against the interestingness-only baseline, it receives stronger ratings for insight-generation support and decision support.}
\end{abstract}




\end{frontmatter}



\section{Introduction}


  
    
   

Recent advances in large language models (LLMs) have increased interest in natural language interfaces (NLIs, \eg, ChatGPT) in both academia and industry.
These interfaces~\cite{cox2001multi,sun2010articulate,gao2015datatone,dhamdhere2017analyza, yu2019flowsense,setlur2016eviza, hoque2017applying, fast2018iris} have made database exploration far more accessible by translating natural language (NL) questions into executable queries (\eg, SQL) and returning appropriate visualizations that support quick insights.
This enables data analysts to analyze data without complex query language or tedious GUI operations (\eg, drag-and-drop).

\virev{However, even when analysts have a broad analytical objective, such as understanding customer buying behavior, they face a critical challenge in \textbf{decomposing that objective into effective step-by-step queries} across unfamiliar multi-table databases.}
The exploration space is large and under-constrained. Analysts must identify which attributes and relationships matter, decide on appropriate aggregations and groupings, and compose a sequence of queries that builds toward their goal rather than producing trivial or fragmented results.
This \emph{operational decomposition} challenge is especially pronounced for analysts who understand their domain but lack familiarity with the specific database schema.
Without structured support, exploratory analysis can easily become shallow and inefficient, leading to potential oversights in critical data insights.


\rev{Existing NLIs~\cite{gao2015datatone, setlur2016eviza, hoque2017applying, narechania2020nl4dv} assist with \emph{executing} user-specified queries, such as translating natural language to SQL, handling ambiguity, and generating visualizations. But, they provide little guidance on \emph{what to explore next}. 
Recent systems have introduced recommendation capabilities with two main approaches, yet with significant limitations.
Interestingness-based methods~\cite{srinivasan2021snowy, lux2021} suggest queries based on statistical patterns in unexplored data subsets, but these recommendations often overlook semantic relatedness and coherence with users' analytical context.
Data-driven approaches~\cite{milo2018next} mine query logs to suggest next actions based on schema- or value-based metrics (\eg, keyword or prefix matching on column names) that do not capture domain semantics or handle cross-table attribute relationships inherent in relational databases.
}


\rev{In this work, we investigate how to augment an NLI with \emph{next-step query recommendations} that are both semantic and context aware and that \mrev{help analysts formulate concrete queries and maintain stepwise context continuity during exploration}.
The recommendations blend schema-level semantic signals, data interestingness, and contextual relevance for relational SQL databases.
Our approach centers on three key aspects: First, we enhance schema semantics by leveraging LLMs to enrich database attributes with concise domain-specific descriptions, enabling more meaningful connections between attributes.
Second, we perform semantic relevance analysis
\mrev{to retrieve semantically similar reference domains and surface recurring query patterns and attribute combinations} that inform candidate recommendations.
Third, we build adaptive contextualization by combining semantic similarity, data interestingness, and a recency-weighted similarity measure to prioritize unexplored and interesting subsets locally aligned with the user's
current analytical focus.
\virev{The resulting recommendations are presented as suggestions that users can inspect, select, refine, or ignore, preserving control over the analytical process.}
To further support NLI-based visual data analysis, the system executes recommended queries, generates appropriate visualizations, maintains an exploration history, and supports lightweight dashboard creation for organizing and communicating insights.}

\virev{We evaluate \systemname{} with NL2SQL benchmarking, schema-description validation, an agentic comparison against two baselines, and a user study with 12 participants.
The agentic evaluation shows broader coverage and higher adoption than the interestingness-only baseline.
In the agentic evaluation, our system receives higher LLM-as-judge ratings than both baselines for topical relevance, context awareness, next-step decision support, discovery, and guidance.
In the user study against the interestingness-only baseline, participants rate it higher for decision support and support for insight generation.
We also derive design implications for future query recommendation systems.}
\virev{Overall, we position \systemname{} as a steerable assistant for analysts who may already have a general objective in mind but need help translating it into executable exploration steps.}
In summary, our major contributions are as follows:
\begin{compactitem}
    \item \textbf{Query recommendation:} A data-driven, semantic- and -context-aware model that suggests next-step NL queries for multi-table SQL analysis by balancing semantic relevance with data interestingness and by exploiting both current and reference query logs. Accordingly, we implement an interactive NLI, \systemname{}, which provides sequential exploration guidance, helping users decide what to ask---while executing queries and visualizing and reviewing results.

    \item \textbf{Evaluation:} Our evaluation, including an agentic technical comparison and a user study with 12 participants, shows improvements in coverage, stepwise coherence, adoption, and perceived usefulness over baseline methods. We also provide design lessons for future systems.
\end{compactitem}





\section{Related Work}
\label{sec.related-work}
Our work builds upon prior research on natural language interfaces and query recommendations for data analysis.

\subsection{Natural Language Interfaces for Data Analysis}
Here, we review the NLIs built for database queries and data visualization.

\textbf{NLIs for database queries}. Many systems have been developed to enable users to access relational databases through natural language, which can be classified into keyword-based systems~\cite{simitsis2008precis}, parsing-based systems~\cite{saha2016athena, kaufmann2006querix, li2014constructing, li2014nalir}, and neural-based systems~\cite{rubin2020smbop, zhong2017seq2sql, guo2019towards, wang2019rat, bogin2019representing, xu2017sqlnet, yu2018syntaxsqlnet}. Recently, large language models (LLMs) have achieved unprecedented performance in natural language understanding and generation. Many researchers use them to translate natural language queries into machine-readable representations like SQL~\cite{xie2022unifiedskg, cheng2022binding}. Furthermore, LLMs can act as user-friendly interfaces~\cite{ma-etal-2023-insightpilot, achiam2023gpt,xie2023openagents} to provide analytical assistance through multi-turn conversations. In our work, we utilize GPT models as the backbone to convert users' queries into SQL for data retrieval.

\textbf{NLIs for data visualization}. 
Many systems combine natural language processing with visualization generation to help users understand data insights. NLIs like Articulate~\cite{sun2010articulate} map NL queries to analytical tasks and decide proper visual encodings, while DataTone~\cite{gao2015datatone} detects and presents data ambiguities in users' NL queries. Given the ambiguity and under-specification of users' queries, many NLIs improve query interpretation via interactive widgets~\cite{gao2015datatone}, semantic parsing pipelines~\cite{narechania2020nl4dv}, and conversation flow characterization~\cite{dhamdhere2017analyza, setlur2016eviza, hoque2017applying, fast2018iris}. However, these NLIs only analyze existing queries and require users to manually craft queries.

Prior work improves NLIs to disambiguate queries and connects them to visual results through ambiguity widgets, semantic parsing, and conversational interaction~\cite{gao2015datatone, yu2019flowsense, setlur2016eviza, hoque2017applying, narechania2020nl4dv, feng2023xnli}. 
These systems primarily interpret current user utterances and produce matching data charts; 
they do not propose future analysis steps. One of the closest works is Snowy~\cite{srinivasan2021snowy}, which recommends next-utterances based on ``interestingness'' of under-explored subsets. However, it is limited to single-table CSVs and pre-defined statistical metrics, making it hard to generalize to multi-table SQL analysis or adapt to domain semantics. 
We target multi-table databases and blend semantic relevance with data interestingness to guide exploration.

\subsection{Query Recommendation for Data Analysis}\label{sec.query-rec-data-analysis}
To reduce manual effort in data analysis, query recommendation techniques assist users in deciding next-step exploration actions. These can be categorized into two main approaches~\cite{milo2020automating}:

\textbf{Interestingness-based systems} evaluate the interestingness of data insights generated by different exploration actions using objective measures (\eg, information gain) or subjective criteria (\eg, unexpected values)~\cite{geng2006interestingness}. 
The typical recommendations include grouping (\eg, roll-up~\cite{sathe2001intelligent}, drill-down~\cite{joglekar2017interactive}), attribute-value pairs~\cite{drosou2013ymaldb}, data charts~\cite{vartak2015seedb}, and data cubes~\cite{sarawagi1998discovery, sarawagi2000user}. However, interestingness measures cannot adapt well to various user preferences.

In contrast,
\textbf{data-driven systems}~\cite{milo2018next, aligon2015collaborative, eirinaki2013querie, chatzopoulou2009query} recommend more personalized next-step actions based on prior queries of the current user or other users, assuming that users with similar query requests may be interested in similar data aspects. 
Data-driven approaches involve two major steps. First, given a user's query contexts, the approaches retrieve the most similar query sequences from query logs generated by the same user or other users.
Then, it analyzes the retrieved sequences to synthesize the final recommendations of next-step exploration for the current user.
However, they often rely on shallow attribute prefix matching and overlook attribute semantics and evolving context.
We bridge these strands by (i) encoding domain semantics and attribute relationships, (ii) retrieving cross-domain reference logs, and (iii) re-ranking candidates by contextual relevance and data interestingness, yielding domain-aligned, context-aware query suggestions.

Complementary to our goal, \textbf{visualization recommendation systems} 
\rev{support chart generation based on pre-defined statistical properties~\cite{vartak2015seedb, key2012vizdeck, demiralp2017foresight}, guided faceted exploration~\cite{wongsuphasawat2015voyager, wongsuphasawat2017voyager}, 
or sometimes including analysis context through design knowledge and chart similarity~\cite{lin2020dziban}.
However, these systems and methods do not explicitly consider NL interaction, data semantics, and relatedness, and their connections with user NL queries. Here, we use standard visualization mappings to display query results. 
\virev{Some work explored personalized visualization recommendation based on user profiles and interaction histories~\cite{qian2022personalized,hu2025interactive}.
They primarily address what to visualize rather than guiding analysts through coherent, multi-step exploratory queries over relational data.}
Our novelty lies in recommending the next semantically relevant, domain-related, context-coherent NL query in multi-table SQL settings.
}

\virev{\subsection{Large Language Models (LLMs) for Data Analysis}\label{sec.llm4da}
Recent systems leverage LLMs to support different aspects of data analysis. InsightLens~\cite{weng2025insightlens} focuses on organizing and exploring conversation histories and insights. LEVA~\cite{zhao2024leva} enhances onboarding, exploration, and summarization with LLMs. LightVA~\cite{zhao2024lightva} and Talk2Data~\cite{guo2024talk2data} emphasize agent-based task planning and question decomposition based on higher-level user goals. The analytical guidance provided by these systems is either retrospective, chart-related, or goal-conditioned. \mrev{Our system complements this by providing incremental scaffolding that helps analysts with a broad domain interest choose concrete next-step queries.} Industrial products like code agents (\eg, Claude Code~\cite{anthropic_claude_code_2026}, Codex~\cite{openai_codex_2026}), ChatGPT~\cite{openai_chatgpt_2026}, Microsoft Power BI Copilot~\cite{microsoft_powerbi_copilot_2025} and Tableau Agent~\cite{tableau_tableau_agent_2026} support language-driven end-to-end analytics workflow. Tableau Agent can suggest data visualizations and insights based on user data. Similarly, code agents can suggest insights autonomously by generating and executing codes. However, these systems generally work as a black box; it is unclear how suggestions are generated. In this paper, we explicitly model and integrate data semantics, data interestingness, and query history into contextual query recommendations. In addition, they tend to prescribe the complete answers instead of suggesting incremental next steps, which limits user ability to steer the analysis or build a coherent mental model of the data.}

\begin{figure*}[ht]
\centering
  \includegraphics[width=1\textwidth]{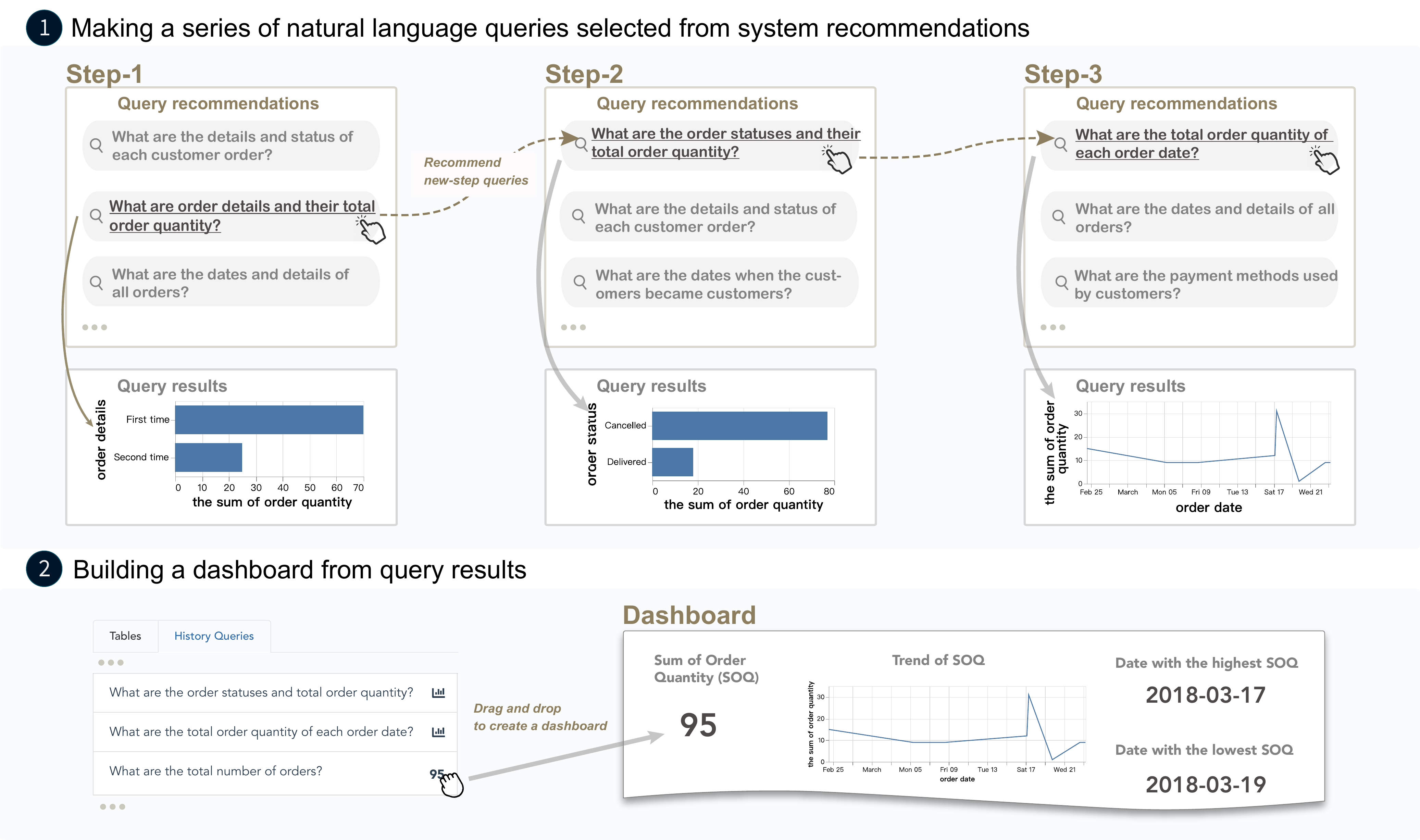}
  \vspace{-5mm}
  \caption{The workflow of using \systemname{} for interactive data analysis. (1) A user first makes a series of natural language data queries by selecting from system recommendations step by step. The results of data queries are presented in visualizations.
  (2) After several steps of data exploration, the user chooses some desired insights (shown in visualizations) from previous queries, and organizes them into a dashboard by direct manipulation.
  }
  \label{fig:teaser}
  \vspace{-5mm}
\end{figure*}

\begin{figure}[ht]
\centering
  \includegraphics[width=1\columnwidth]{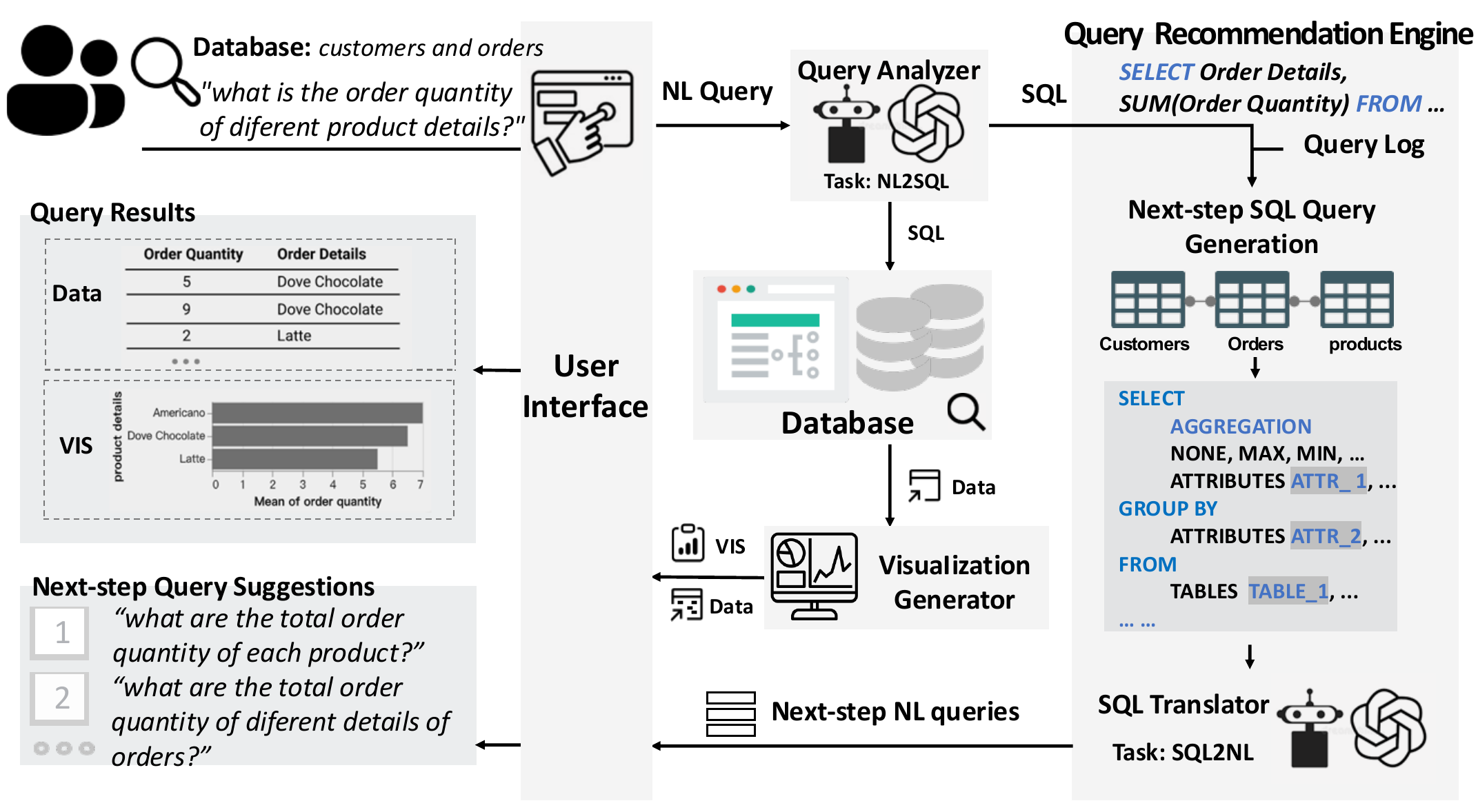}
  \vspace{-5mm}
  \caption{
  System framework. After a user submits a NL query of a specific domain through the \ui{}, the \nlsql{} will translate it into SQL queries to retrieve relevant data from databases, and the \vg{} will automatically visualize the data.
  Meanwhile, based on the prior queries made by the current user in the \qlog{} and the \textit{Reference Queries} from databases, the \rec{} generates next-step natural language exploration queries for the \textit{Target Database Tables}. 
  These recommendations are converted into NL forms to promote the understanding of system responses.
  Eventually, the data and visualization of the current query and next-step NL query suggestions will be presented in the \ui{}.}
  \label{fig:system_framework}
\end{figure}
\section{Design Requirements}
\label{sec.design_req}

\virev{Our target users are data analysts who may have a general analytical objective but need operational support in navigating unfamiliar database schemas, especially those with limited experience in multi-table SQL exploration.}
\rev{We aim to develop an NLI that recommends stepwise data exploration actions to data analysts to help them decide step-by-step queries when exploring data insights in unfamiliar databases.}
To identify the design requirements of \name{}, we first reviewed the design requirements and implementations of previous NLIs for data analysis~\cite{cox2001multi,sun2010articulate,gao2015datatone,dhamdhere2017analyza, yu2019flowsense,setlur2016eviza, hoque2017applying, fast2018iris, narechania2020nl4dv,li2014constructing,li2014nalir,srinivasan2021snowy,feng2023xnli}.
\rev{In addition, we collected feedback from external participants and one internal design collaborator to refine the design requirements.}

\rev{\par \textbf{University students} (\imp{S1-S5}, 3 males 2 females; aged 19-34) with backgrounds in computer science (S1, S2), statistics (S3, S4), and information systems (S5).
They rated themselves as beginner or intermediate users of data visualization tools and programming for data analysis. All had prior experience with NLIs such as ChatGPT~\cite{openai_chatgpt_2026}.}
\par \rev{\textbf{Industry practitioners and internal collaborator} (\imp{D1-D4}, four males; 3 aged 25-34, 1 aged 35-44): three external data analysts (\imp{D1-D3}) and one internal data visualization scientist (\imp{D4}) from an international technology company.
\imp{D1} (8 years experience, database engineer), \imp{D2} (6 years experience, cloud computing analytics), and \imp{D3} (5 years experience, algorithm engineer) regularly use BI tools such as Tableau and Microsoft Excel, and write code for data analysis across different business domains (\eg, sales and cloud computing). 
\imp{D4} (6 years experience, a co-author of this paper) specializes in data visualization research and tool development. 
}

\rev{Across interviews, participants consistently identified that
even when they had a general analytical direction, decomposing it into effective queries over unfamiliar schemas was a major struggle.
As \imp{S1} and \imp{S3} articulated, \textit{``I don't know how to start my first question (\imp{S3}).''}
\textit{``I am always worried that my questions do not cover some important data aspects, and things I've found are boring''} (\imp{S1}).
\imp{D3} specifically requested that the system should \textit{``present meaningful exploration actions that show which attributes are relevant for the investigated domains,''} rather than simply translate NL to SQL.
\virev{The core challenge is not a lack of analytical intent but the difficulty of mapping that intent to concrete queries over unfamiliar table structures and attribute relationships.}
Recommendations should support the analysts \textit{``without disrupting the analysis flow \imp{(D1)}''} and \textit{``should not replace human judgement \imp{(S2)}''}. 
}

\xbRevvv{During the design process, we carried out weekly meetings with data analysts and the internal design collaborator for about seven months and with students for about four months, and iteratively updated the design requirements and system prototypes according to their feedback.
For example, after developing an NLI with the recommendation module, a summary of explored queries and data visualizations was desired and added to promote data provenance and organize data insights.
Finally, we consolidated a list of design requirements as follows.
}


\textbf{R1. Provide easy access to databases via natural language queries.}
Natural language interaction provides an intuitive and user-friendly way for users to interact with databases and conduct the analysis flow~\cite{fast2018iris, dhamdhere2017analyza, setlur2016eviza, hoque2017applying}. 
Our target users also desire to use NL to quickly formulate their data needs.
In addition, to promote data discovery,
NLIs need to offer hints on what data exists in databases and what queries systems support~\cite{setlur2016eviza, yu2019flowsense}.
\imp{D3} recommended supporting 
autocompletion of data attributes when typing a question, helping compile their questions.


\textbf{R2. Recommend next-step exploration actions that operationalize users' analytical objectives.}
\virev{Data analysts often have a broad analytical goal (\eg, ``understand customer buying behavior'') but need support in decomposing that goal into concrete, executable data queries across unfamiliar multi-table schemas.}
However, \imp{D1} and \imp{D3} pointed out that
they often need to spend lots of time finding interesting domain-specific data facts when the dataset is from a domain that they are not very familiar with.
\virev{For instance, a domain-specific fact could be that a particular product has disproportionately high order quantities for an e-commerce dataset, or which department has an unusually low student-to-faculty ratio for an education dataset.}
Moreover, the large data exploration space and high data complexity in databases make data exploration even more challenging~\cite{milo2018next, aligon2015collaborative}.
To reduce manual effort in insight discovery, \imp{D3} suggested that the system should present meaningful next-step exploration actions (\example{which attributes are relevant for the investigated domains}).
In addition, data analysis is a subjective and iterative process that involves multiple exploration steps.
Analysts may have diverging analysis flows,
i.e., their analytical interests can also change during exploration.
Thus, the system is expected to offer context-aware recommendations that are \mrev{locally coherent with the user's prior exploration} and  dynamically adapted to users' analytical focuses~\cite{sordoni2015hierarchical,milo2020automating,srinivasan2021snowy}.

\xbRevvv{\textbf{R3. Balance semantic relevance and data interestingness for query recommendation.}
Since the typically large and complex nature of datasets, prior work~\cite{milo2018next,srinivasan2021snowy} has explored metrics for identifying data subsets with interesting patterns (``data interestingness'').
However, those subsets may not be semantically relevant to users' analytical contexts implied in NL queries. 
\imp{S2} stated, \textit{``interesting patterns are not helpful if they don't relate to my question.''}
This mismatch can cause confusion and distrust~\cite{zehrung2021vis} about \textit{``why the system recommends such queries \imp{(S5)}.''}
Therefore, the recommendation should prioritize data subsets that contain interesting patterns and semantically align with query contexts to facilitate coherent data exploration.
}

\textbf{R4. Explain the relevance of system responses to users' queries.}
We find that not all our target users have expertise in database query languages (\eg, SQL) and visualization.
Thus, the system should explain analysis operations powered by SQL (\eg, attribute selection) and generate visualizations in an understandable manner.
\imp{D4} said that it would be better if
the system could use NL to present suggestions on exploration actions (\ie, SQL-related operations).
\imp{D3} added that the system needs to demonstrate the mappings between retrieved data and generated visualizations.
Besides, 
\imp{D3} stated that the system should link input NL queries with the generated SQL, allowing him to verify if it retrieves the correct data.

\textbf{R5. Support an easy revisit of previous queries and results.}
Data analysis is a multi-step iterative process.
Many of our target users mentioned that they often need to refer back to previous queries, review what insights they derive, and adjust the future exploration path.
Afterward, the analysts need to select important data facts and the corresponding visualizations from prior queries and create a complete data story using a dashboard.
\imp{D1} advised saving and summarizing user query sequences and restore prior queries on demand. \imp{D2} recommended organizing the queries and visualizations of interest into a dashboard.

\section{\systemname{}}


Motivated by the derived requirements, we design and implement \name{} (\autoref{fig:teaser}) that can recommend next-step exploration actions to facilitate NL-based interactive visual data analysis.


\subsection{System Framework}
\xb{\autoref{fig:system_framework} summarizes the system workflow. 
After loading a database of interest, users can either manually input an NL query or select one from the system recommendations. 
The \nlsql{} translates the input query into a SQL query using LLMs
and retrieves the data from databases. The \vg{} automatically generates visualizations based on the data's properties. 
Meanwhile, the \rec{} proposes next-step queries, balancing semantic relevance, user context, and data interestingness based on users' query logs. The recommendations cover attribute selection, aggregation, grouping, and filtering.
\virev{The suggested SQL queries are then translated into NL for readability in the User Interface. To help users check potentially ambiguous interpretations, such as underspecified intents or similarly named attributes across tables, the system also presents the retrieved data for inspection and query reformulation.}
}


\subsection{Next-step Exploration Recommendations}
\label{subsec:exploration_rec}

\begin{figure}[!htbp]
\centering
  \includegraphics[width=.9\textwidth]{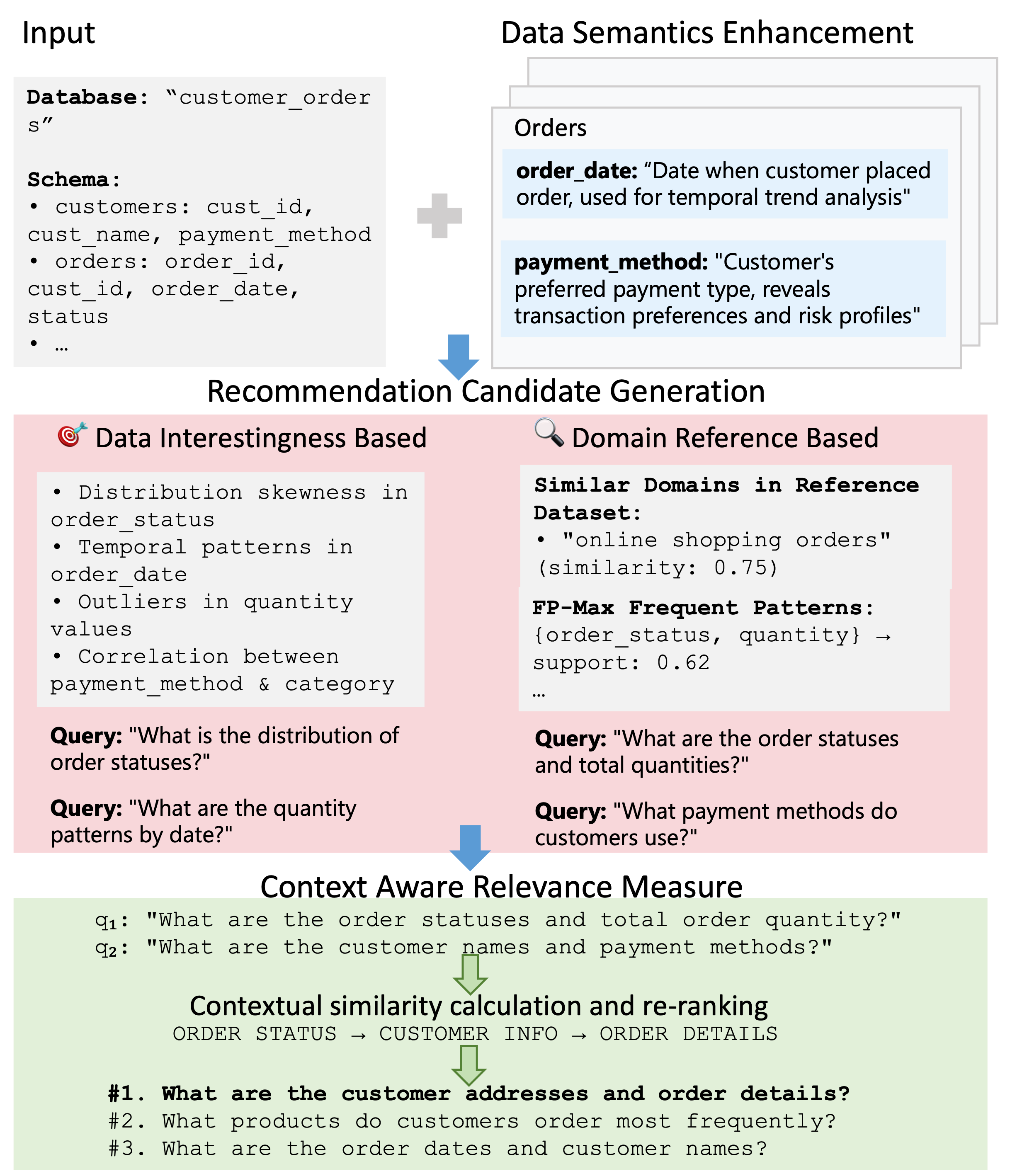}
  \caption{\xb{The pipeline of query recommendation. First, data semantics are enhanced for data attributes of tables in the target database. Then, recommendation candidates are generated using data interestingness- and domain reference-based methods. Finally, the candidates are ranked using contextual similarity measures.}}
  \label{fig:rec-pipeline}
\end{figure}

We propose a data-driven method (\autoref{fig:rec-pipeline}) to generate semantically relevant and context-aware next-step exploration actions. Our approach balances semantic relevance (how well recommendations align with the analytical domain interests) with data interestingness (how likely the data subset has meaningful patterns).

\subsubsection{Problem \& Data}
Given the database $D_{0}$ with target schemas (tables/columns), a short domain tag $Dom_{0}$ (or database name, \eg, customer service), the problem is to generate query suggestions $Q_{s}$ in NL according to the current user's previous queries $Q = \{q_{0}, q_{1}, ..., q_{n}\}$ and, potentially, historical queries from other users, serving as references. The query references come from Spider~\cite{yu2018spider}, an external large-scale dataset containing 10,000 diverse user SQL queries across 138 analysis domains, such as commerce, government, and education, reflecting actual database usage.

\subsubsection{Enhance Data Semantics and Relatedness}\label{subsec.data-semantics-enhance}
\mrev{Data attributes are a central source of domain knowledge, which is valuable for domain insight discovery.}
\mrev{However, these attributes can be brief or cryptic, limiting our ability to understand their semantic relationships.}
To address this, we enhance data semantics in two aspects.

\textbf{Semantic enhancement of attributes}. We leverage LLMs (GPT-4o) to expand column names with concise, schema-grounded explanations.
These expanded explanations enable better grouping of related attributes and more contextually relevant recommendations.
\begin{exampleblock}{1}
\begin{tabularx}{\linewidth}{@{} l X @{}}
\textbf{Input}: &
Table=\code{orders}, Column=\code{order\_date} \\
\textbf{Output}: &
The date an order was placed, can be used for sales trend analysis.
\end{tabularx}
\end{exampleblock}

\textbf{Semantic relevance analysis.}
After semantic enhancement of data attributes, we leverage users' query logs to incorporate more explicit user interests and implicit domain knowledge.
Specifically, when a user selects a target domain (\eg, customer order deliveries), our system retrieves similar domains (\eg, customer order addresses) from the reference dataset to see what analysts typically asked in a similar context.
We use sentence transformers (all-MiniLM-L6-v2) for transforming domains (database names) into embeddings to balance speed and performance for semantic search via cosine similarity computation.
The resulting reference users' queries reveal different attribute combinations that help the system to formulate meaningful and relevant queries for the user.

\subsubsection{Initial Exploration Action Recommendation}\label{subsec.initial-rec}
\mrev{Before a user submits an initial query, our model generates SQL data query suggestions.}
If no reference domains are retrieved for the target database,
the system will compute data interestingness metrics (\eg, unevenness, deviation, skewness) based on Lux~\cite{lux2021} to derive top-ranked data subsets for exploration.

\rev{
\textbf{Data interestingness metrics} 
compute the statistically notable data patterns based on data types (See 
\autoref{tab:interestingness} in Appendix for more details).  
For categorical (discrete) data, \textit{unevenness} captures distribution imbalance across categories, while \textit{chi-square} tests independence between categories.
For quantitative data, skewness detects asymmetric distributions and monotonicity for trend and correlation analysis.
Metrics are sorted in descending order of magnitude.
}

\textbf{\mrev{Semantic-based recommendation}.}
If reference domains are available, the corresponding user query logs are used to compose the SQL query in the form of \texttt{SELECT <attributes> <AGG optional> FROM <tables/join> [GROUP BY \\<attributes (optional)>]} through three key steps.
The process begins with \textit{finding frequent attributes}. For each column in the target database, we compute its semantic similarity with columns in reference queries. 
If a column's enhanced semantics match many columns in reference queries above a similarity threshold, it likely interests the current user. For example, in a retail database, attributes like ``product\_name,'' ``order\_quantity,'' and ``customer\_id'' frequently appear in reference queries, indicating their analytical importance.
Next, we \textit{discover attribute combinations} using the FP-max algorithm~\cite{grahne2003efficiently} to find frequent co-occurrence patterns among attributes. 
For example, if attributes matching ``order\_quantity'' and ``order\_status'' often appear together in reference queries, we recommend them simultaneously.
During this process, we record which SQL clauses and operations these attributes participate in. For instance, ``order\_status'' might be used in SELECT clauses while also serving as a GROUP BY attribute, leading to a query that groups orders by \texttt{order\_status} and counts the rows.
Finally, we \textit{rank by data interestingness}. 
After generating SQL candidates, we execute them and measure data interestingness using Lux. We then rank recommendations to prioritize those with interesting patterns.

\subsubsection{Context-Aware Exploration Action Recommendation}
\mrev{As users progress through their analysis, our system generates query candidates and ranks them against recent query history so that recommendations remain locally coherent with the user's current exploration context.}

\textbf{Query candidate generation.} We consider two complementary strategies for generating candidates. 
First, we identify frequent attribute combinations and their corresponding SQL clauses in the target database that involve at least one unexplored attribute, leveraging cached FP-Max results for efficiency. Second, we compute unexplored data subsets with the highest data interestingness scores to surface potentially valuable patterns.

\textbf{Contextual relevance measure.}
\mrev{Based on these candidates, we employ a contextual similarity measure with recency decay to re-rank candidates by local semantic relevance and recent-context continuity:}
\begin{equation}
C_{\mathrm{sim}}(c)=\sum_{i=0}^{n}\alpha^{n-i} R_a(c,q_i),
\qquad 0 < \alpha < 1,
\end{equation}
where $c$ is a candidate query, $q_i$ is the $i$-th prior user query, and $q_n$ is the most recent prior query. The decay term $\alpha^{n-i}$ gives the most recent query weight 1 and progressively discounts older queries\footnote{\mrev{We empirically set $\alpha = 0.7$ to balance recent-query continuity and broader session context.}}. $R_a(c,q_i)$ denotes the attribute-level relevance between candidate $c$ and prior query $q_i$, linearly combining similarity to matched reference-query attributes and to the user's previous query attributes. Specifically, the similarity is computed by the maximum paired cosine similarity between attributes in query candidates and the attributes in the corresponding query logs.
In this way, if a user starts by exploring ``customer names'' and then asks about ``customer payment methods'', the system recognizes the focus on customer-related analysis and prioritizes recommendations like:
``What are the customer contact details?'', ``When did customers become customers?'', ``What are the customer billing addresses?''.

\mrev{\textbf{Component Summary.} Together, these components target complementary recommendation qualities: schema-description enrichment and reference-domain retrieval support semantic relevance, frequent attribute mining provides candidate structures from related domains, data interestingness prioritizes statistically notable subsets, and recency-weighted contextual ranking maintains continuity with the user's recent analytical focus.}

\xbNote{\textbf{Translation between NL and SQL}. After coming up with the final recommendations of data exploration actions, we convert the SQL data queries back to natural language to help users understand. We leverage OpenAI GPT-3.5-turbo for translation purposes, considering the balance among task performance, model performance, inference speed, and cost. 

\rev{To translate NL to SQL for the user-specified database exploration, the system first extracts all tables, columns, and primary-foreign key relationships by reading the database schema.
Then, the schema together with a user query are injected into LLM prompt template as context to generate a SQL query. We leverage GPT-4o for NL-to-SQL translation, where the formal evaluation is in \autoref{sec:nl2sql-eval}. The prompts of translations are included in \autoref{sec.prompt_llm_nl2sql}.}
}

\subsection{Data Visualization}
\label{subsec:data_vis}
After retrieving data from databases, \name{} generates data visualizations to show the results. The visualization designs are automatically generated according to the type of data. 
For example, the data of the answer to the query, \example{the total order quantity of each customer}, are pairs of the total order quantity (\textbf{Q}uantitative) and customer (\textbf{N}ominal). This type of data (\textbf{Q}$\times$\textbf{N}) is visualized in a bar chart.
We adopt the framework from NL4DV~\cite{narechania2020nl4dv} to map different types of data to a list of visualization candidates from a wide range of common visualization types (\eg, bar chart, line chart, and scatter plot). 
The visualizations are formulated in vega-lite \cite{satyanarayan2016vega}, a high-level visualization language, and rendered in the user interface. 
For the cases when the data is too complicated and no proper visualization is found to show the results, \name{} presents the raw data in a table. When the data simply contains one element (\eg, \example{the total order quantity}), \name{} directly displays the data value.

\begin{figure*}[!htbp]
\centering
  \includegraphics[width=.9\textwidth]{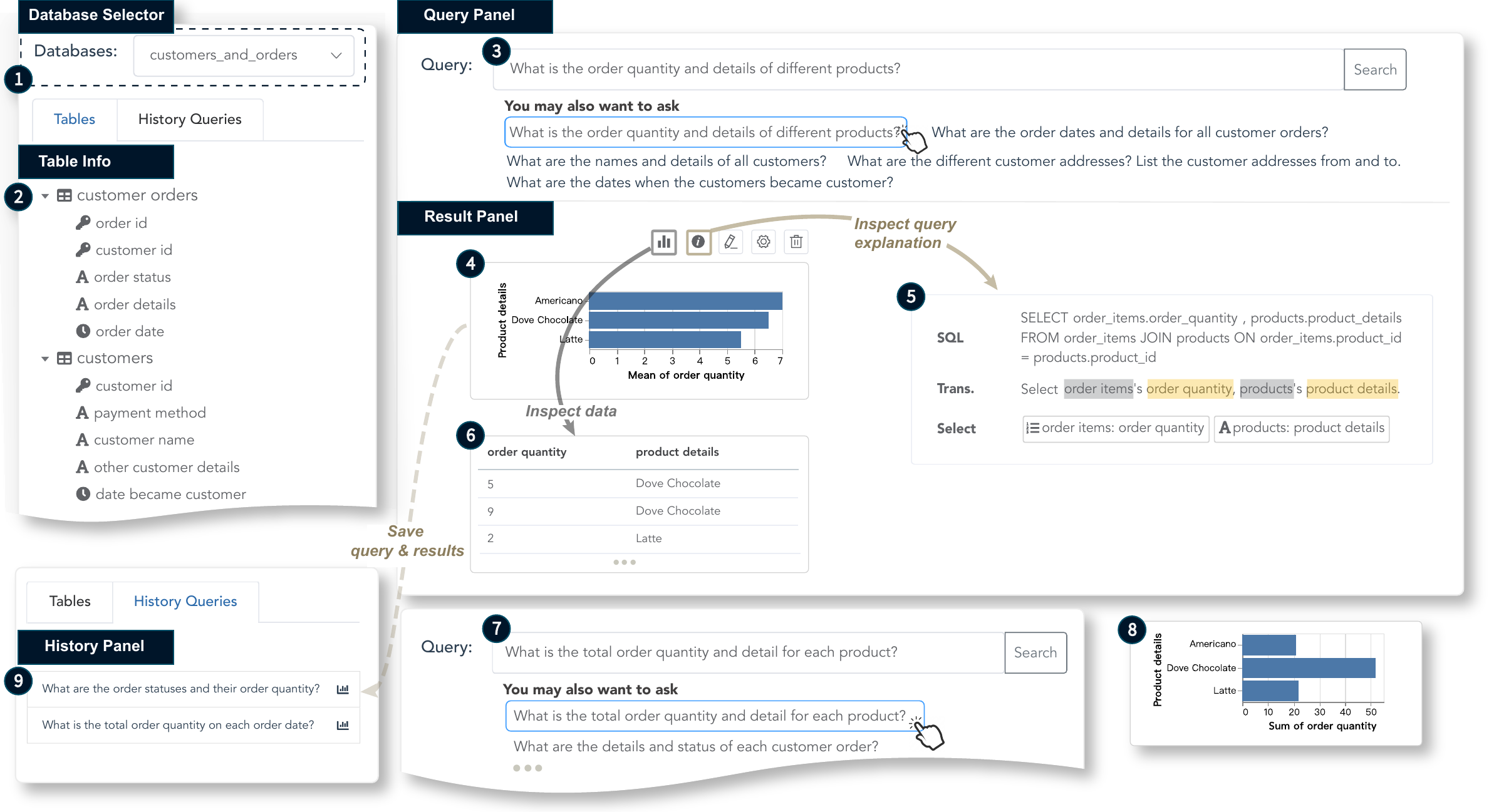}
  \caption{A system walkthrough.
  A user first selects a target database in the \dbs{} (1) and checks the database details within the \ti{} (2). 
  After making the initial query by selecting the first system recommendation in the \qp{} (3)
  he inspects the query results (4) in the \rp{}, checks the template-based explanations of SQL components (5) to the query processing procedure, and the retrieved data in a table (6).
  He confirms that the system interprets the natural language query correctly. Then he continues to make further explorations with more queries (7, 8). And his previous queries are stored in the \hp{} (9).
  }
  \label{fig:system_walkthrough}
\end{figure*}

\subsection{System Walkthrough}
\label{subsec:sys_walk}
Based on the recommendation model, we build an NLI for data analysts to perform interactive data analysis with step-wise exploration guidance and automatic data visualizations. 
Moreover, users can create a dashboard to summarize the data insights generated during the exploration process.

In this section, we introduce our system workflow, visual components, and interaction designs through a usage scenario where Andy, a data analyst from a sales department,
uses \name{} to perform NL-powered interactive data analysis.
Andy is tasked to analyze data about \emph{customer behavior} in retail stores.
In addition, he needs to create a dashboard to inform store managers of the extracted insights into the selling situation and customer characteristics.
Although Andy knows about context information about customers and products,
he has limited knowledge of SQL and store sales data. He then refers to our system to interactively explore data insights through NL.

\subsubsection{Start Initial Exploration}
Andy first chooses to explore a customer-oriented sales
database in the \dbs{}, named \example{customers\_and\_orders} (\autoref{fig:system_walkthrough}.1).
After loading the database, he gains an overview of the data tables and columns of the databases in the \ti{} (\autoref{fig:system_walkthrough}.2).
Afterward, he shifts his attention to the recommended queries in the \qp{} before he starts the initial query.
He finds the top five recommendations (\autoref{fig:system_walkthrough}.3) are all quite reasonable and clicks the first one \example{What is the order quantity and details of different products?} to discover the distribution of product order quantity.

\subsubsection{NL2SQL Explanation \& Data Visualization}
After submitting the query, Andy examines the corresponding system response (\autoref{fig:system_walkthrough}.4) in the \rp{}.
To ensure that \name{} retrieves the correct data from databases, he first refers to the predicted SQL and its explanations (\autoref{fig:system_walkthrough}.5) by clicking on~{\faInfoCircle}~in the resulting visualization.
After reading the highlighted terms (\example{\textcolor{ent}{\textbf{order items}}, \textcolor{col}{order quantity}}, \example{\textcolor{ent}{\textbf{products}}, \textcolor{col}{product details}}) in the NL explanation,
he realizes that the system retrieves the values of attribute \textcolor{col}{order quantity} from the table \textcolor{ent}{order items} and the values of attribute \textcolor{col}{product details} from the table \texttt{\textcolor{ent}{products}}.
Furthermore, he inspects the retrieved data of his query by clicking on~{\faBarChart}~in the visualization.
Finally, he confirms that the system interprets his query correctly.
Then, Andy examines the generated data visualization (\autoref{fig:system_walkthrough}.4).
The bar chart describes how many orders for each product there are on average.
He notices that \example{Americano} has the largest average number of orders (with the longest bar) while \example{Latte} has the least average number of orders (with the shortest bars).
He would like to investigate more about product orders.

\subsubsection{Stepwise Exploration with Query Recommendations}
He refers to query recommendations (\autoref{fig:system_walkthrough}.7) to know what he can ask about product orders in the next query.
Among the suggested queries,
he chooses the first one 
\example{What is the total order quantity and details for each product?}
The resulting bar chart below (\autoref{fig:system_walkthrough}.8) shows the total orders for each product, where \example{Dove Chocolate} has the most orders (with the longest bar).
Afterward, he is also interested in analyzing other dimensions of product orders.
And he iteratively refers to the recommendations, formulates a series of queries
about product orders (\eg, \example{order details}, \example{order status}, \example{order date})~(\autoref{fig:teaser}.1).
Meanwhile, he inspects the corresponding visualizations to derive the data insights.
For example,  a majority of products are ordered for the first time rather than the second time,
most product orders are canceled rather than delivered, and a large increase and decrease of product orders are observed between March 17 and March 21.

\subsubsection{Dashboard Construction}
After several exploration iterations with NL, 
Andy feels ready to create a dashboard to summarize the data insights. 
He examines his previous queries in the \hp{} (\autoref{fig:system_walkthrough}.9) and clicks them to restore the query results in the \rp{} of the system.
He then picks several queries and visualizations,
which describe the insights about product orders from different dimensions.
Eventually, he organizes them into a dashboard by direct manipulation (\eg, drags and drops) (\autoref{fig:teaser}.2).
Next, he will use this dashboard to communicate the data insights about product orders with store managers.

\section{Evaluation}
\label{sec.evaluation}

\rev{We evaluate \systemname{} with four complementary aspects:
(1) a \emph{technical evaluation of the NL2SQL module} to assess LLM-generated SQL quality;
(2) a \emph{technical evaluation of LLM-enhanced descriptions} to assess the correctness and stability of our semantic enhancement step;
(3) a comparative \emph{technical evaluation of recommendation quality} characterized by semantic relevance and data interestingness using an LLM agentic framework;
and (4) a \emph{user study} assessing perceived recommendation quality, workflow support, and usability.}

\rev{The interestingness-only baseline used in both the agentic evaluation and user study shares the same natural-language interface, NL-SQL translation modules, visualization generator, history panel, and dashboard construction tools as \systemname{}.
These two interface conditions differ only in the recommendation strategy: the interestingness-only baseline ranks candidates solely by data interestingness,
while \systemname{} incorporates both contextual semantic relevance and data interestingness.
Baseline~II is included only in the agentic technical evaluation and replaces the recommendation engine with direct GPT-4o prompting.}


\subsection{Technical Evaluation of NL2SQL}
\label{sec:nl2sql-eval}
\rev{We evaluate the standalone NL2SQL module used in \name{} on the development set of the Spider dataset (Spider-dev). 
The split contains 1,034 natural language questions across different domains.
It typically involves 1.51 tables and 2.73 columns (see \autoref{tab:spider_dev_complexity} in Appendix).
We evaluate three LLMs, including GPT-4o, GPT-4o-mini, and GPT-3.5-turbo, using our NL2SQL prompting template (\autoref{sec.technial_evaluation}).
We vary the number of in-context demonstrations (few-shot examples) and test 0-shot, 1-shot, and 3-shot settings (see more details in \autoref{sec.technial_evaluation}). 
We report execution accuracy (EX), the standard Spider metric that compares the execution results of the predicted SQL against the ground truths.
To account for randomness in LLM generation, we run each configuration three times and report the mean and standard deviation.
}

\begin{table}[ht]
\centering
\caption{\lsy{Evaluation on Spider-dev under different few-shot settings.}}
\label{tab:spider-selection}
\setlength{\tabcolsep}{9pt}
\renewcommand{\arraystretch}{1.0}
\begin{tabular}{c|c|c|c}
\toprule
\textbf{Few-shot} & \textbf{GPT-4o} & \textbf{GPT-4o-mini} & \textbf{GPT-3.5-turbo} \\
 & \textbf{EX} & \textbf{EX} & \textbf{EX} \\
\midrule
0-shot & 75.3 ($\pm$0.2) & 76.5 ($\pm$0.3) & 74.1 ($\pm$0.3) \\
1-shot & 81.2 ($\pm$0.3) & 77.7 ($\pm$0.4) & 76.4 ($\pm$0.3) \\
3-shot & 78.2 ($\pm$0.1) & 78.3 ($\pm$0.3) & 78.0 ($\pm$0.2) \\
\bottomrule
\end{tabular}
\end{table}

\textbf{Results Analysis.}
\rev{As shown in \autoref{tab:spider-selection}, all three LLMs achieve reasonable and stable performance on Spider-dev, with execution accuracy in the range of 74\%-81\% and low variance across runs.
The best execution-accuracy setting is GPT-4o with 1-shot demonstration (EX = 81.2\%).
The results demonstrate that our lightweight NL2SQL prompt can produce good performance without requiring model finetuning.
We also report some case analyses in the Appendix \autoref{table:nl2sql-error-easy}, \autoref{table:nl2sql-error-medium} and \autoref{table:nl2sql-error-hard}.
}

\subsection{Technical Evaluation of LLM-Enhanced Descriptions}
\label{sec:semantic-desc-eval}

\virev{Because \name{} uses LLM-enhanced column descriptions as the semantic basis for reference-query retrieval and recommendation ranking, we evaluate whether these descriptions are both correct and stable. We conduct two analyses on ten databases spanning diverse domains, covering column-level description items. First, two independent human experts (2M, both over 4 years of experience in data analytics) assess each generated description based on schema-related information, such as the table and column names, column type, and key relations. We use a 3-point factual correctness scale (2 = fully correct, 1 = largely correct with minor extrapolation, 0 = incorrect or unsupported), and explicitly count unsupported business semantics as 0. Second, we assess generation stability by regenerating each description three times and measuring pairwise cosine similarity (based on their embeddings), BERTScore, and ROUGE-L across runs. Detailed protocols, per-database results, and representative failure cases are reported in \autoref{appendix.correctness_description}, \autoref{appendix.stability_description}.}

\textbf{Results Analysis.}
\virev{The human evaluation shows that the generated descriptions are generally correct and well grounded in schema evidence. Averaging across both experts, the mean factual-correctness score is 1.75 out of 2; the individual expert means are 1.63 and 1.87, with Spearman's $\rho = 0.61$ ($p < 0.001$), indicating reliable judgments. The stability analysis further shows high semantic consistency across repeated generations, with overall cosine similarity of 0.862 and BERTScore of 0.832, while ROUGE-L is lower at 0.634. This gap suggests that most cross-run variation is lexical rather than semantic. Taken together, these results suggest that LLM-enhanced descriptions provide useful and generally reproducible schema-grounded semantic signals for recommendation.}

\subsection{Technical Evaluation of Recommendation Quality}
\label{sec.tech-eval}

The effectiveness of \name{} largely depends on whether our system can generate high-quality query recommendations to guide visual data analysis, especially assessing the integration of data interestingness and semantic relevance.

\virev{\textbf{Baselines.} We compare our methods against two baselines. 
\textbf{Baseline I} (\emph{interestingness only}), commonly used in prior work~\cite{srinivasan2021snowy, lux2021}, greedily chooses queries for unexplored data subsets in SQL tables with the largest interestingness scores.
The implementation is based on Lux~\footnote{\url{https://github.com/lux-org/lux}}. 
The comparison assesses the added value of \emph{query semantic relevance and coherence}.
\textbf{Baseline~II (LLM-Direct)} is a prompt-based approach to assess whether the benefit of a structured pipeline of query recommendation can be achieved by LLMs directly. Given the full database schema and the user's query history, we prompt GPT-4o to suggest next-step natural language queries that are semantically relevant, diverse, and analytically coherent directly. \lsyRev{The full prompt is provided in the supplement \autoref{app:baseline-ii-prompt}.}
}


First, we sample 8 databases from the Spider dataset to cover various domains, such as e-commerce, education, software, and entertainment. 
\rev{For each dataset-system pair, an LLM-based Data Analyst Agent conducts a multi-turn exploratory analysis session.
Following existing evaluation protocols that cap interaction rounds to a small fixed budget~\cite{wang2023rethinking}, we use six agent actions per session: one cold-start query followed by five next-step decisions.
\mrev{We analyze the resulting query sequences and retrieved data using:
(i) automated coverage metrics;
(ii) agent adoption of recommendations; and (iii) a rubric-based LLM-as-judge assessment of recommendation quality, including semantic relevance, next-step guidance, and perceived support for insight discovery.}
}


\begin{figure*}[!htbp]
\centering
  \includegraphics[width=1.0\columnwidth]{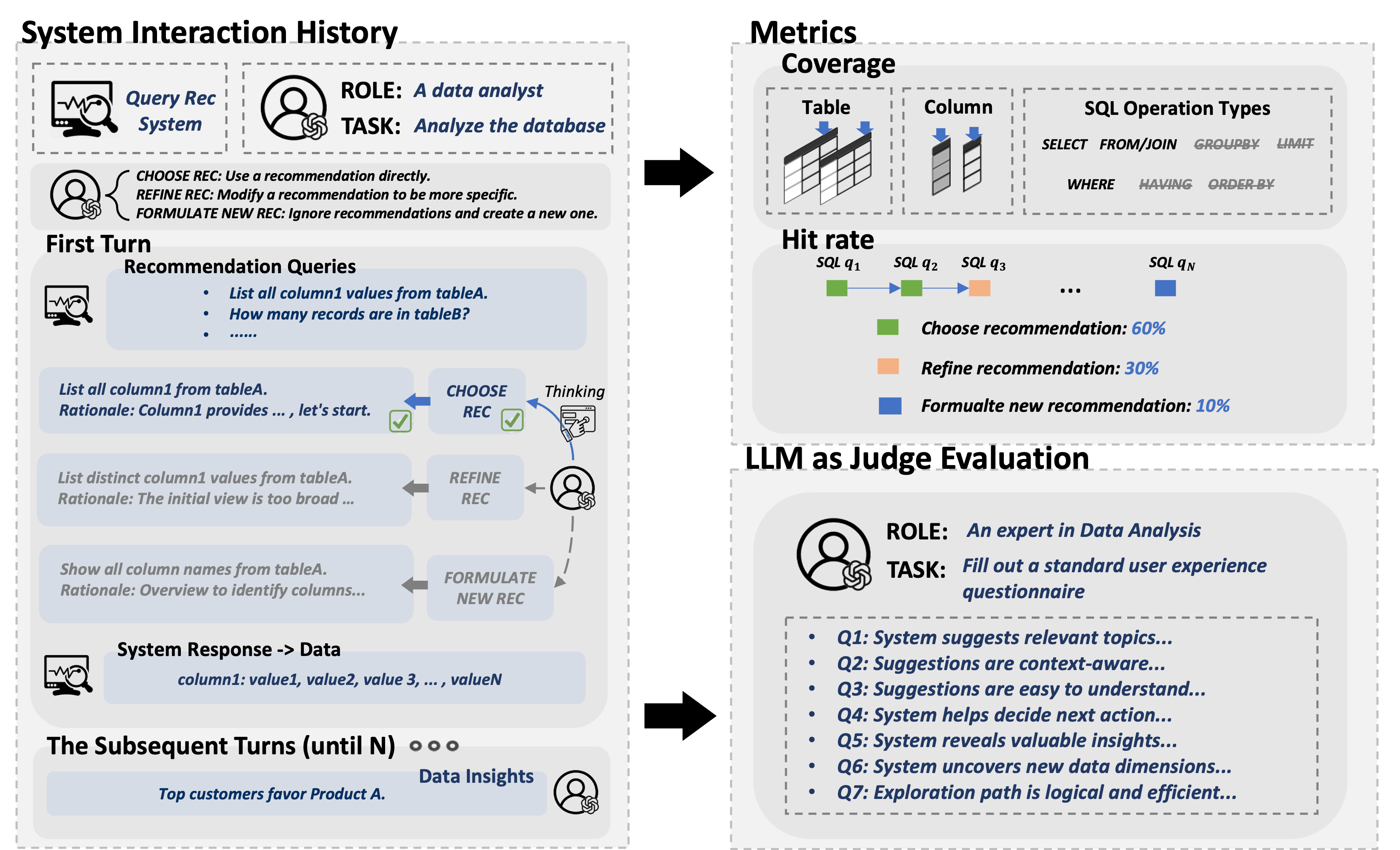}
  \caption{The agentic system design for automated technical evaluation. We leverage an LLM (\ie, GPT-4o) to simulate data analysts to interact with all three query recommendation systems (\ie, Baseline~I, Baseline~II, and \systemname{}) in multiple turns. After simulation, based on the query logs, we compute coverage and hit rate and leverage another LLM (\ie, GPT-4o) as a judge to evaluate the data exploration query logs.}
  \label{fig:eval_pipline}
\end{figure*}

\subsubsection{Metrics and Agentic Evaluation}
\label{sec.tech-metrics}
As shown in \autoref{fig:eval_pipline}, our agentic pipeline evaluates a recommendation system on a given database by producing a multi-turn exploration trace. Then, we score it quantitatively and qualitatively.

\textbf{Data analyst agent} (simulation). Motivated by the cost of repeated human-in-the-loop trials and the strong role-play/analysis capabilities of modern LLMs, we use an LLM agent (GPT-4o) to simulate analysts. 
The agent maintains memory of prior queries and results and, at each turn, chooses among three actions given the system’s current recommendations:
\begin{itemize}
    \item CHOOSE a recommendation verbatim;
    \item REFINE a recommendation (e.g., add GROUP BY, filters); or
    \item FORMULATE NEW if recommendations are off-topic.
\end{itemize}
The agent outputs its rationale and the final query at each decision point. After the six-action session, we compute all metrics on the resulting trace.

\textbf{Evaluation aspects.}
\lsy{\autoref{table:llm-as-judge-aspects} provides detailed definitions of all automated evaluation aspects.}
\rev{The coverage and hit rate directly quantify how recommendations shape the \emph{exploration breadth} and the extent to which suggestions are directly adopted. LLM-as-judge rubrics evaluate the \emph{semantic qualities} such as topical relevance, contextual awareness, next-step guidance, and perceived impact on insight discovery.}

\emph{Coverage metrics.}
Given the simulation results, we quantify exploration breadth using four coverage measures (see \autoref{table:llm-as-judge-aspects}), including table, column, aggregation, and SQL clause coverage.
We compare methods per database using Wilcoxon signed-rank tests.

\emph{Hit rate} is reported for recommendation adoption rate: the proportion of agent actions where the agent chooses a system recommendation (vs. refine or formulate new).

\emph{LLM-as-judge} (quality \& impact). A separate LLM judge (GPT-4o) rates session logs on a six-item, five-point rubric (Q1--Q6): topical relevance, context awareness, clarity, next-step decision support, discovery of underexplored attributes or analytical dimensions, and overall guidance (prompts in the supplement~\autoref{sec.prompt_llm_eval}).

\begin{table*}[h]
\centering
\footnotesize
\caption{\lsy{Definitions of evaluation metrics, including coverage, LLM-as-judge, and hit rate. All metrics are higher-is-better.}}
\label{table:llm-as-judge-aspects}
\setlength{\tabcolsep}{0.2pt}
\renewcommand{\arraystretch}{1.0}
\begin{tabular}{p{0.26\linewidth} p{0.74\linewidth}}
\toprule
\textbf{Aspect} & \textbf{Definition} \\
\midrule

\textbf{Table Coverage} & 
The proportion of unique tables referenced by the system’s recommended SQL queries relative to all tables in the database schema. \\

\textbf{Column Coverage} & 
The proportion of unique columns referenced by the system’s recommended SQL queries relative to all columns in the database schema. \\

\textbf{Aggregation Coverage} &
The extent to which the recommended SQL queries utilize the set of standard aggregation functions (COUNT, SUM, AVG, MAX, MIN). \\

\textbf{Clause Coverage} &
The extent to which the recommended SQL queries include diverse SQL clauses (e.g., WHERE, JOIN, GROUP BY, ORDER BY, LIMIT). \\

\midrule
\textbf{Relevance} & The system generates topically relevant suggestions for the user's domain of interest. \\

\textbf{Context-awareness} & The system provides context-aware suggestions for the user's next exploration step. \\

\textbf{Clarity} & The user can easily understand the natural language query suggestions recommended by the system. \\

\textbf{Next Step Decision} & The system effectively helps the user decide on the next exploration action. \\


\textbf{Discovery} & Through its recommendations, the system helps the user discover previously underexplored data attributes or analytical dimensions. \\

\textbf{Guidance} & Guided by the system, the user's exploration path is logically coherent, strategically efficient, and successfully achieves its core analytical goals. \\

\midrule
\textbf{Hit Rate} & 
The proportion of all agent actions that directly select one of the system’s recommended queries (CHOOSE\_RECOMMENDATION), reflecting how often the system’s recommendations are adopted in the simulated analysis process. \\

\bottomrule
\end{tabular}
\end{table*}



\subsubsection{Result Analysis}
\label{sec.tech-eval-result}

\begin{figure*}[ht]
    \centering
    \includegraphics[width=1\linewidth]{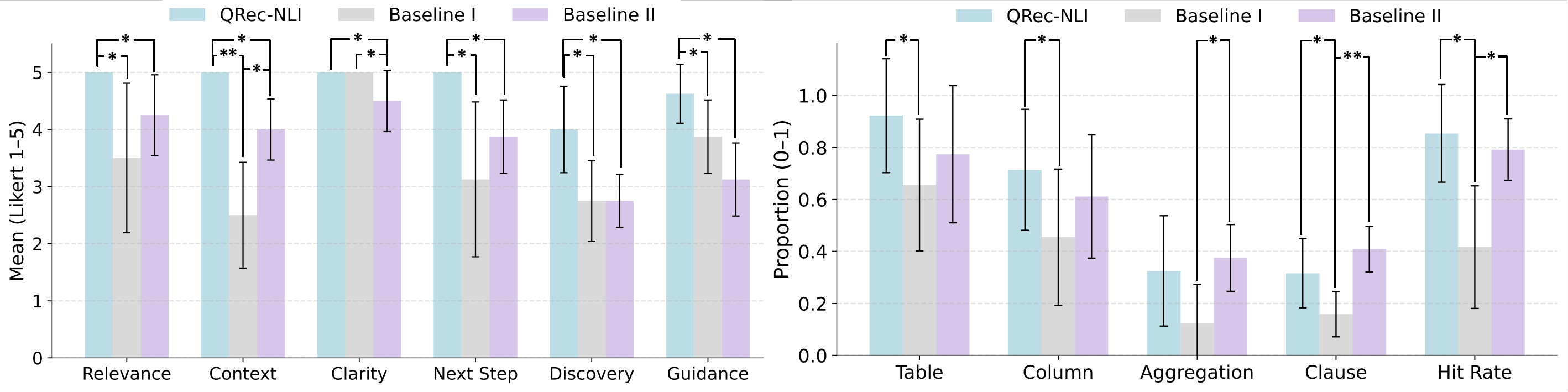}
    \caption{\lsy{Means and standard deviations of automated LLM-as-judge scores comparing QRec-NLI with Baseline~I (Interestingness-only) and Baseline~II (LLM-Direct) (*: $p<.05$, **: $p<.01$).}}
    \label{fig:llm-as-judge}
\end{figure*}

\textbf{Quantitative results.} 
\autoref{fig:llm-as-judge} presents the comprehensive evaluation metrics across 8 Spider databases (detailed breakdowns are in \autoref{table:detailed-ratings}, \autoref{table:coverage-comparison-full}, and \autoref{table:user-interaction-types}).
\mrev{As a validation check, we compared GPT-4o judge ratings with two student raters on \systemname{} session logs and observed strong ordinal agreement across all three raters (Krippendorff's $\alpha_{ord}=0.83$; see \autoref{appendix.human-llm-align}). Because this check uses student raters and \systemname{} logs rather than domain analysts across the full comparative setting, we interpret LLM-as-judge scores as supporting comparative evidence of recommendation quality, not definitive evidence of real insight quality or domain decision quality.}
\rev{The agentic evaluation results are analyzed using Wilcoxon signed-rank tests.}
\systemname{} consistently outperforms Baseline~I on exploration breadth, recommendation quality, and recommendation adoption.
Relative to Baseline~I, \name{} significantly improves table coverage (92\% vs. 65\%, W = 1.0, p = 0.046), column coverage (71\% vs. 46\%, W = 0.0, p = 0.018), and clause coverage (31\% vs. 16\%, W = 0.0, p = 0.026), although aggregation coverage is not statistically significant (W = 6.0, p = 0.085).
The LLM-as-judge ratings also favor \systemname{} on relevance (5.00 vs. 3.50, W = 0.0, p = 0.026), context-awareness (5.00 vs. 2.50, W = 0.0, p = 0.011), next-step guidance (5.00 vs. 3.12, W = 0.0, p = 0.017), discovery (4.00 vs. 2.75, W = 0.0, p = 0.026), and overall guidance (4.62 vs. 3.88, W = 0.0, p = 0.034), while both systems achieve equally strong clarity. \systemname{} also yields a significantly higher hit rate (85.42\% vs. 41.67\%, W = 0.0, p = 0.018).

\par \virev{Against the stronger prompt-based Baseline~II, \systemname{} still achieves the highest LLM-as-judge average score on all eight databases (4.77 vs. 3.75) and significantly improves relevance (5.00 vs. 4.25, W = 0.0, p = 0.034), context-awareness (5.00 vs. 4.00, W = 0.0, p = 0.011), clarity (5.00 vs. 4.50, W = 0.0, p = 0.046), next-step guidance (5.00 vs. 3.88, W = 0.0, p = 0.014), discovery (4.00 vs. 2.75, W = 0.0, p = 0.015), and overall guidance (4.62 vs. 3.13, W = 0.0, p = 0.016). In contrast, the differences between \systemname{} and Baseline~II are not statistically significant on table coverage, column coverage, aggregation coverage, clause coverage, or hit rate (all $p > 0.05$). \mrev{Overall, under this agentic evaluation, direct prompting is a stronger baseline than interestingness-only ranking, yet the structured pipeline receives higher LLM-as-judge ratings for semantic coherence and next-step guidance.}}


\textbf{Case analysis.}
\xb{\systemname{} demonstrates topical relevance with suggestions like ``show all customer names'' and then ``count customers'',
that are directly aligned with understanding customer information. 
The system shows strong context awareness, fostering exploration coherence between queries.
For example, one \systemname{} session presents the following logical progression: (1) exploring customer names, (2) quantifying total customers, (3) examining customer status, (4) analyzing payment methods, and (5) investigating contact channel completeness.
This systematic approach helps discover meaningful insights, such as all 15 customers sharing the same ``Standard'' status and data completeness issues in contact information. In contrast, the interestingness-only baseline's path is more fragmented, with the agent noting that ``static suggestions did not fully support a logical flow, requiring the user to drive the exploration.''
The impact on discovery is significant. 
\lsyRev{\systemname{} better supports the discovery of previously unknown data attributes or analytical dimensions through more diverse recommendations, such as payment method distributions and potential data quality issues.}
The interestingness-only baseline, despite finding some insights, tends toward more repetitive information and requires more user initiative to formulate meaningful queries when recommendations are too basic (\eg, repeatedly suggesting to explore empty fields). 
\virev{LLM-based prompting approach more often produces plausible broad follow-up questions, but it still underperforms \systemname{} on contextual continuity and overall guidance, indicating that direct prompting alone is less reliable for sustaining a coherent exploration trajectory.}
}

\subsection{User Study}


\xb{
We conducted a lab study with 12 participants (P1–P12) to (1) elicit feedback on working with query recommendations during exploratory data analysis and (2) assess system design choices and surface areas for improvement. To contextualize feedback, we compared QRec-NLI against the interestingness-only baseline used in \autoref{sec.tech-eval}.
}

We used a within-subjects design with two conditions (\systemname{} vs. the interestingness-only baseline) and two datasets. 
\rev{Both conditions provided the same visual data exploration interface, as described at the beginning of \autoref{sec.evaluation}.
The only difference is in the query recommendation results shown below the query input box.}
The order of systems and datasets was counterbalanced across participants to reduce ordering effects. Participants followed a think-aloud protocol.


\textbf{Tasks and procedures.}
\xb{
Participants were tasked to perform data analysis using the interestingness-only baseline and \name{} in a think-aloud manner. 
}

\xb{Each study session lasted about 70 minutes on average. Initially, we collected participants' demographic information and asked their permission to use their personal data generated during the study for research purposes only. Then, we spent about 15 minutes briefing participants about the background and procedures of our study. 
Next, participants were requested to use \name{} and the interestingness-only baseline 
to explore datasets I and II, respectively, where they also needed
to create dashboards. The order of systems and tasks was counterbalanced. Participants did not know the data details before the study. The time limit for each system was about 15 minutes.
Before each task, we gave a short five-minute tutorial on how to use the system and phrase queries using an example dataset. 
The participants could report their data insights early. 
After each task, they would have another 5 minutes to organize discovered insights into
a dashboard and then were asked to finish a questionnaire \lsy{(see \autoref{table:user-study-aspects})}
to evaluate the system they had just used. 
Eventually, we conducted post-study interviews with the participants, collecting their feedback on the exploration experience of using the two NLI systems.}

\textbf{Data and tasks.}
\xb{We used two databases from Spider that do not require specialized knowledge to explore and are not used in the technical evaluation.
Database \texttt{I} is about ``customers\_and\_addresses'' which contains seven tables, including customer orders, customers, order items, products, addresses, customer addresses, and customer contact channels, and 34 columns in total. The goal is to report some valuable insights to store managers, informing them of the selling situation and customer buying behavior. Database \texttt{II} is on student transcript tracking, containing 11 tables with 49 columns in total. The goal is to act as a teacher to analyze student performance.}
\mrev{Both datasets were chosen to support open-ended, non-specialist exploration. We counterbalanced system order and system-dataset pairing to reduce ordering and assignment effects, but their different domains and schema sizes mean that insight difficulty was mitigated rather than fully matched.}

\textbf{Participants.}
We recruited 12 participants (\imp{P1-P12}), nine males and three females, age 18-34, from local universities.
They came from different academic backgrounds, such as computer science, automation, mathematics, optoelectronics, and sensing. 
They had varying knowledge levels of SQL and chatbot-based data analytics tools. For SQL proficiency, 6 rated themselves as entry-level users, 4 as intermediate users, and 2 as proficient users. For chatbot familiarity with data analytics tools like ChatGPT, 2 were beginners, 2 were regular users, and 8 were proficient users. They frequently use LLMs/AI tools during data work (daily: n=6; weekly: n=5; once or twice: n=1).
All participants had normal vision and hearing, and we compensated each participant with \$16.

\virev{
\textbf{Measures and link to system designs.}
Core recommendation capabilities and interface features are evaluated through a structured questionnaire and user feedback, \virev{guided by validated HCI and recommender system evaluation constructs}.
In the questionnaire (see \autoref{table:user-study-aspects}), four items, adapted from prior evaluation frameworks \cite{pu2011user, srinivasan2021snowy, seo2025mt}, target the effectiveness of the recommendation engine, including \emph{topical relevance}, \emph{context awareness}, \emph{decision support}, and \emph{insight generation}.
These dimensions are directly influenced by contextual semantic enhancement.
\emph{Data retrieval accuracy}, \emph{visualization quality}, \emph{NL comprehension}, \emph{query review}, \emph{organization support} align with the information quality and interface quality subscales of the PSSUQ~\cite{lewis2002psychometric} and Snowy \cite{srinivasan2021snowy}.
\emph{Learnability}, \emph{overall usability} are adapted from System Usability Scale \cite{brooke1996sus}, while \emph{intrusiveness} reflects the non-intrusiveness criterion for proactive recommendation systems~\cite{srinivasan2021snowy, melguizo2007proactive}.
}


\begin{table}[!htbp]
\centering
\small
\caption{Definitions of questionnaire evaluation aspects in the user study.}
\label{table:user-study-aspects}
\setlength{\tabcolsep}{2pt}
\renewcommand{\arraystretch}{1.1}
\begin{tabular}{p{0.33\linewidth} p{0.64\linewidth}}
\toprule
\textbf{Aspect} & \textbf{Definition} \\
\midrule
\textbf{Topical Relevance} & Suggestions were topically relevant to my domain of interest. \\

\textbf{Context Awareness} & Suggestions were context-aware for my next step. \\

\textbf{Decision Support} & The system helped me decide my next step. \\

\textbf{Insight Generation} & The system helped me find reasonable insights. \\

\textbf{Confidence Building} & I felt confident in choosing the next step. \\

\textbf{Data Retrieval} & The system retrieved the right data fields for my queries. \\

\textbf{Visualization Quality} & The visualizations of retrieved data were intuitive. \\

\textbf{NL Comprehension} & The NL suggestions were easy to understand. \\

\textbf{Query Review} & It was easy to review past queries and results. \\

\textbf{Organization Support} & The system helped organize/summarize queries \& results. \\

\textbf{Learnability} & It was easy to learn. \\

\textbf{Usability} & It was easy to use. \\

\textbf{Intrusiveness} & Suggestions were intrusive or distracting. \\
\bottomrule
\end{tabular}
\end{table}

\subsection{User Study Results and Analysis}
\label{sec.user-study-results}
We analyze participants' ratings using Wilcoxon signed rank tests to compare \systemname{} to the interestingness-only baseline, and thematic analysis of user feedback to assess the effectiveness and usability of \systemname{}.



\subsubsection{Quantitative Outcomes}
\xbNote{
The results 
demonstrated significant improvements in several key areas of the recommendation system.
\systemname{} significantly outperformed the baseline in core recommendation quality metrics.
For \textbf{topical relevance}, \systemname{} achieved a mean rating of 4.42 (SD = 0.67) versus 3.33 (SD = 1.07) for the baseline, W = 4, p = 0.016.
Similarly, \textbf{context awareness} showed significant improvement (\systemname{}: M = 4.33, SD = 0.49; Baseline: M = 3.58, SD = 0.79; W = 0, p = 0.031).
The system also demonstrated significant advantages in supporting users' analytical processes. 
\textbf{Decision support} was rated significantly higher for \systemname{} (M = 4.58, SD = 0.51) than the interestingness-only baseline (M = 3.58, SD = 0.90), W = 3, p = 0.023. 
\textbf{Insight generation} similarly favored \systemname{} (M = 4.58, SD = 0.51) over the interestingness-only baseline (M = 3.67, SD = 1.07), W = 4, p = 0.031.
While not statistically significant, \systemname{} showed positive trends in other important areas. Confidence building improved from 3.50 (SD = 1.31) in the interestingness-only baseline to 4.50 (SD = 0.80) in \systemname{} (p = 0.055), approaching significance. 
Data retrieval accuracy also showed improvement (\systemname{}: M = 4.42, SD = 0.79; Baseline: M = 3.50, SD = 1.09; p = 0.063).

Both systems were perceived to have comparably good usability. 
Visualization quality, NL comprehension, query review capabilities, and organization support all received ratings above 4.0 for \systemname{}, with no significant differences from the interestingness-only baseline. 
Both systems were rated as easy to learn (\systemname{}: M = 4.67, SD = 0.65; Baseline: M = 4.42, SD = 1.16) and easy to use (\systemname{}: M = 4.50, SD = 0.67; Baseline: M = 4.17, SD = 1.34).
Notably, both systems maintained low intrusiveness levels (\systemname{}: M = 2.08, SD = 1.31; Baseline: M = 2.17, SD = 1.03), indicating that \mrev{the recommendation features did not disrupt users' analytical workflows.}
}

\subsubsection{Semantic Understanding and Contextual Awareness}
\xb{Participants consistently praised \systemname{}'s semantic understanding and ability to stay on-topic compared with the interestingness-only baseline.
When exploring customer orders, P6 noted that \systemname{} \textit{``naturally transitioned to recommend order items--exactly my exploration plan''}, demonstrating the system's ability to anticipate logical analytical progressions. 
P3 observed that after asking about course credits, the system \textit{``also suggested combining course name and grade, which matched my goal''}, indicating effective data context awareness and coherence. 
Users also appreciated temporal semantic reasoning.
P6 liked how \systemname{} \textit{``understood 'recent months' as a relative time range and showed a trend chart''.} 
In contrast, interestingness-only baseline suggestions were often characterized as \textbf{``broad, lacking context awareness, semantically incoherent''} (P3) and exhibiting \textbf{``off-topic drift'' }(P4). 
P11 noted that interestingness-only baseline recommendations were \textbf{``redundant and possibly misleading,''} with some complex queries returning no data.
These qualitative observations strongly support our quantitative findings on topical relevance and context awareness.}

\subsubsection{Impact on Analytical Workflow and Insight Discovery}
\systemname{} enhanced users' ability to discover meaningful patterns in their data. 
Participants reported that the system directly guided them toward valuable findings that might otherwise be overlooked. 
P1 specifically noted the system flagged \textit{``product\_id = 2''} when analyzing order quantity. Through the chart, he found that the product in order \#10 had 14 items, a concrete outlier (a popular product), which he could have missed without the recommendation. 
P6 similarly discovered that \textit{``product\_id = 2 had the highest sales, with order \#10 containing 14 items and order \#7 containing 9 items,''} leading to inventory management insights about maintaining \rev{over 20} units of safety stock for this high-demand product.

The recommendations were perceived as particularly valuable in two critical situations:
At the very beginning, to establish a sensible starting point, and immediately after an initial probe, to continue the line of queries.
As P3 put it, recommendations ``help most after I introduce a factor'', frequently serving as bridges from coarse overviews to targeted follow-ups.

The history panel and dashboard construction helped participants maintain orientation and consolidate findings.
P4 described the history panel as ``crucial--like my mind map. I can jump back anytime and avoid getting lost'', while P2 noted that it ``avoids repeated work'' and helps ``shift from single points to overall trends''.
P11 noted similar benefits: ``History helps me record past queries, preventing redundant questions.''
Participants favored query-management interactions supported by \systemname{}, such as quick query revisits and result restores, and dashboard organization.

\xbNote{Summary. These quantitative and qualitative results confirm that \systemname{}'s semantic and context-aware approach significantly enhances the core aspects of query recommendation---topical relevance, contextual adaptation, decision support, and insight discovery--while maintaining good usability comparable to the interestingness-only baseline.}

\subsubsection{User Adaptation Strategies and System Limitations}
When system recommendations did not meet users' expectations, participants showed some adaptive behaviors that reveal both system strengths and limitations.
Some users intentionally ignored breadth-oriented suggestions when pursuing focused analysis.
P4 explained, ``I follow a depth-first strategy...the system's suggestions were breadth-first, so they conflicted with my current strategy''. 
This selective engagement showed that users maintained control by skipping irrelevant suggestions.

However, participants also raised some issues.
Cold-start breadth occasionally lacked depth. P2 mentioned that suggestions would ``only list fields...lack analytical direction'', prompting requests for trend/compare-type prompts.
Some complex or loosely related attribute combinations yielded empty results, especially in the interestingness-only baseline or when both systems shifted topics prematurely. 
Users either rephrased or reverted back to previous queries via \hp{}.
The interestingness-only baseline particularly struggled with aggregation (\eg, ``all scores together'' instead of by term/course), providing ``only raw details without aggregation; trends are not obvious'' (P6). 
Participants therefore needed to manually inject sorting and grouping operations.
Column naming quirks (e.g., spaces) occasionally required rewording to ``hit'' the correct fields (P11); participants learned to rephrase the queries.


\section{Discussion}
Based on the design and evaluation of \name{}, we discuss the design implementations for query recommendations, 
critical factors for user trust and adoption, 
automation, agency, and adaptiveness for data analysis, and system limitations and future work.

\subsection{Design Implications for Query Recommendations}
Grounded in our quantitative and qualitative evaluation, we derive the following design implications for query recommendations.

\xbNote{
\textbf{Manage query complexity at proper scope and granularity.}
During the user study, some participants observed that the system sometimes generated overly complex queries with many attributes or filtering conditions.
These queries led to overwhelming data tables that users could not easily interpret, or to empty results that confused them.
This suggests that query recommendations should adapt query complexity to the underlying data characteristics. For larger data subsets, the system should further decompose exploration into smaller and more manageable sets.
For sparse or empty data, the system can consider revealing the data transformation process and avoiding over-filtering. 

\textbf{Support non-linear exploration and reflection.}
Our current history panel supports users revisiting previous queries and results, as well as the recommendations at the moment, in a linear manner. 
Some participants suggested branching exploration paths, allowing them to pursue multiple analytical threads without losing context. 
In addition, future systems can support undo operations to handle errors or misleading suggestions (P4), along with comparison and preview mechanisms to help users evaluate recommendations before committing to them (P3).
 
Moreover, participants alternated between breadth-first and depth-first strategies. Systems can provide clear mechanisms for switching between overview, comparison, and detailed analysis modes and receiving appropriately scoped recommendations.

\textbf{Be data quality aware and break the low-value loop.}
An unexpected finding from our LLM-based simulation revealed that systems can become trapped in repetitive data quality validation loops, repeatedly checking for nulls or data consistency rather than advancing analytical goals. This suggests that recommendation systems need explicit data quality models that perform initial validation once, then adapt recommendations based on known data characteristics rather than repeatedly rediscovering the same quality issues. The system can also detect repetitive, low-signal cycles and propose a goal-aligned alternative.
}

\subsection{Automation, Agency, and Reliance in Interactive Data Analysis}
Interactive data analysis is a challenging task that requires collaborative work between humans and AI.
Ultimately, effective query recommendation requires a delicate balance between intelligent automation and user agency. 
On the one hand, the system must be smart enough to provide valuable guidance while remaining aligned with the user's intent. 
Here, we leverage the user query logs and semantic relevance to identify interesting insights and adapt to the user's analytical interest.
\mrev{On the other hand, user agency should be respected and defended because helpful recommendations can also create over-reliance, especially when users are unfamiliar with the database schema.
In \systemname{}, a recommendation may look reasonable because it follows the recent analysis context, but it may still be based on partial signals, such as attribute-level similarity, reference-query patterns that do not cover all relevant alternatives, or ambiguous matches between a natural-language query and database fields.
Therefore, users should not be encouraged to accept suggestions uncritically.
They need lightweight ways to inspect why a query is suggested, preview the retrieved data, refine or reject the suggestion, and redirect the analysis when it no longer matches their intent.
Future evaluations should also measure reliance more directly, such as whether recommendations reduce independent query formulation or verification behavior.
These considerations suggest that future query recommendation systems should move beyond simple pattern matching or statistical interestingness toward context-aware, steerable systems that treat data exploration as a collaborative dialogue between human intent and machine intelligence.}

\subsection{User Confidence in Query Recommendations}
Our user study revealed that user confidence related to multiple factors:
(1) predictability of recommendation result; 
(2) clear explanation of why a recommendation is relevant;
(3) alignment with users' domain knowledge and current tasks;
and (4) successful query execution.

Our current system leverages semantic relevance and promotes context coherence, which helps users stay on topic, build a mental model of system capabilities, and track analytical progression.
In contrast, for the baseline system, as P4 observed: ``unexplained jumps—recommending unrelated tables or fields without reasons—reduce confidence.'' 
Nevertheless, when unexpected recommendations occur, the system should provide light-weight evidence, such as 
attaching a short one-line rationale referencing recent action, 
display decision confidence and simple provenance (like source data table included in Results panel),
and using some social contexts, \eg, ``show something like 85\% adoption by similar users.''
Conversely, queries returning empty results, query errors, and repetitive surface-level suggestions prevent users from adopting system suggestions.
The future systems can consider some safety nets around recommendations, 
such as data profiling (\eg, missingness) and give users warnings that inform query designs, or explain failure modes and suggest robust transforms.

\subsection{Limitations and Future Work}

\textbf{Enable more data operations.}
Our system recommends three basic types of exploration actions for data analysis, including attribute selection, data grouping, and aggregation.
More complex data operations need to be considered to enhance the system utility in the future,
for example, arithmetic operations (\ie, multiplication, addition, subtraction, and division) of attribute values.
In addition, our query recommendation techniques consider the semantics of SQL queries. Other language features (\eg, ontology) can be utilized to characterize data queries.

\textbf{Build guardrails for query recommendations.} 
\name{} recommends next-step queries by considering the semantic relationships among dataset headers, attributes, and users' queries.
\virev{In practice, however, real-world databases can be noisy, underspecified, and ambiguous, arising from multiple valid join paths, similarly named attributes across tables, or underspecified analytical intents. 
To support users in detecting ambiguity-related errors, our system provides post-hoc SQL inspection mechanisms by exposing SQL explanations and retrieved data for user inspection. However, this safeguard is reactive. 
Users must notice the problem and reformulate the query themselves. Future work should support schema curation for noisy headers and add proactive ambiguity handling, such as detecting competing join paths or attribute bindings before execution, surfacing alternative interpretations, and enabling lightweight clarification or editable correction interactions~\cite{gao2015datatone, narechania2021diy, elgohary2020speak}.}

\mrev{\textbf{Capture complementary but semantically distinct tasks.} Our contextual similarity measure operates at the attribute level, using LLM-enriched descriptions and reference-domain co-occurrence patterns to support local semantic relevance and stepwise continuity. It may not capture a full analytical plan. When useful next steps involve semantically distant attributes or patterns absent from reference logs, the method may fail to connect them as complementary parts of a broader objective. Future work could incorporate task-level intent representations or LLM-based planning to reason over such broader objectives.}

\mrev{\textbf{Component-level ablations.} Our evaluation compares QRec-NLI with the interestingness-only and LLM-Direct baselines, but does not isolate individual modules. Further ablations on schema enrichment, reference-domain retrieval, frequent-pattern mining, interestingness ranking, and contextual reranking can reveal their effects on relevance, discovery, and multi-step coherence.}

\mrev{\textbf{Enhance evaluation.} Our LLM-as-judge evaluation supports comparative interpretation of recommendation quality in a controlled simulated setting, but it is not definitive evidence of real insight quality or domain decision quality. The user study counterbalanced system order and dataset assignment, but the two datasets were not fully matched for domain familiarity or intrinsic insight difficulty. Therefore, we interpret the user-study results as evidence of perceived recommendation quality and workflow support, rather than as a fully controlled comparison of insight difficulty.}

\section{Conclusion}
To facilitate interactive visual data analysis, \systemname{} augmented NLIs with stepwise, semantically relevant, context-aware query recommendations for multi-table SQL databases by leveraging users' query logs. \virev{It helps analysts operationalize broad analytical goals into inspectable query steps while preserving user control.}
It executes queries with visualizations and supports history review and dashboard construction.
\mrev{Across a four-part evaluation, \systemname{} shows stable NL2SQL and schema-description performance. In the agentic comparison, it achieves broader coverage and higher adoption than the interestingness-only baseline and receives higher LLM-as-judge ratings than both baselines for topical relevance, context awareness, next-step decision support, discovery, and guidance. In the user study against the interestingness-only baseline, participants rate it higher for decision support and insight-generation support.}

\section*{Acknowledgement}
The project was funded by Nanjing University - China Mobile Communications Group Co., Ltd. Joint Institute (No. NJ20250035).

\section*{GenAI Usage}
During the preparation of this work, the authors used ChatGPT and Claude to polish English word choice, sentence composition, and grammar. The authors reviewed and edited the content as needed and take full responsibility for the published article.





\bibliographystyle{elsarticle-num}
\bibliography{main}

\appendix
\renewcommand{\theHfigure}{\Alph{section}.\arabic{figure}}
\renewcommand{\theHtable}{\Alph{section}.\arabic{table}}
\renewcommand{\theHequation}{\Alph{section}.\arabic{equation}}
\section*{Supplementary Materials}

\lsy{This section provides additional details that complement the main paper, including definitions of the data interestingness metrics, full prompt templates, extended quantitative results, and error analyses. All terminology and notation follow the definitions in the main text so that the supplementary material can be read independently when needed.}

\section{Data Interestingness Metrics}
\label{sec.data-interestingness}
\rev{This section describes the computation of data interesting metrics (in \autoref{tab:interestingness}), including unevenness, skewness, deviation, monotonicity, and Chi-square.}

\paragraph{Unevenness (categorical distribution).}
Unevenness quantifies the degree of imbalance in a categorical distribution.
For a categorical attribute with cardinality $C$ and normalized frequency vector
$\mathbf{v} = [v_1, v_2, \ldots, v_C]$ where $\sum_{i=1}^{C} v_i = 1$,
unevenness is computed as:
\begin{equation}
    \text{Unevenness}(\mathbf{v}) = D \cdot \|\mathbf{v} - \mathbf{v}_{\text{flat}}\|_2
\end{equation}
where $\mathbf{v}_{\text{flat}} = [1/C, 1/C, \ldots, 1/C]$ represents a uniform distribution,
$\|\cdot\|_2$ denotes the Euclidean ($L_2$) distance, and $D = 0.9^C$ is a cardinality
penalty factor that discounts distributions with many categories to account for
reduced visualization readability~\cite{mackinlay1986automating}.
Higher unevenness scores indicate more distributional imbalance.

\begin{table*}[t]
\centering
\caption{\rev{Data interestingness metrics used for evaluating candidate data subsets.
Metrics are chosen according to data patterns and visualization types. M: measure, D: dimension, T: time attributes.}}
\resizebox{\linewidth}{!}{
\begin{tabular}{l l l l}
\toprule
\textbf{Data Pattern} & \textbf{Visualization Type} & \textbf{Metric} & \textbf{Interpretation} \\
\midrule
Categorical distribution & Bar chart (1D+1M) & Unevenness &
Higher scores indicate dominant or rare categories \\
Value distribution & Histogram (1M) & Skewness &
Detects asymmetric distributions or outliers \\
Comparative deviation & Filtered view (subset vs.\ overall) & Deviation &
Measures how a filtered subset differs from the overall dataset \\
Correlation & Scatter plot (2M, 1T+1M) & Monotonicity &
Detects linear or monotonic trends \\
Categorical independence & Colored bar / heatmap (2D+1M) & Chi-square &
Tests if categories are independent \\
\bottomrule
\end{tabular}}
\label{tab:interestingness}
\end{table*}

\paragraph{Skewness (value distribution).}
Skewness quantifies the asymmetry of a numerical distribution around its mean.
For a quantitative attribute $X$ with mean $\mu$ and standard deviation $\sigma$,
skewness is defined as the standardized third moment:
\begin{equation}
    \text{Skewness}(X) = \frac{\mu_3}{\sigma^3} = \frac{\mathbb{E}[(X - \mu)^3]}{\sigma^3}
\end{equation}
where $\mu_3 = \mathbb{E}[(X - \mu)^3]$ is the third central moment.
Positive skewness ($\text{Skewness} > 0$) indicates a distribution with an asymmetric
tail extending toward higher values (right-skewed), while negative skewness
($\text{Skewness} < 0$) indicates a tail extending toward lower values (left-skewed).

\paragraph{Deviation (subset vs.\ overall).}
Deviation measures how much a filtered subset $S \subseteq D$ differs from the overall
dataset $D$.
For a categorical or binned quantitative attribute with normalized frequency vectors
$\mathbf{v}_D$ (overall) and $\mathbf{v}_S$ (filtered subset), deviation is computed as:
\begin{equation}
    \text{Deviation}(S, D) = \text{sig}(S) \cdot \text{rankSig}(S, D) \cdot \|\mathbf{v}_D - \mathbf{v}_S\|_2
\end{equation}
where:
\begin{itemize}
    \item $\text{sig}(S) = |S| / |D|$ is the \textit{filter significance} factor,
          representing the proportion of data retained. Larger filtered subsets
          receive higher weight as they are more representative and statistically stable.
    \item $\text{rankSig}(S, D)$ is a \textit{rank change factor} for ordered categorical data,
          defined as:
          \begin{equation}
              \text{rankSig}(S, D) = 1 + \frac{1}{C} \sum_{i=1}^{C} |\text{rank}_D(c_i) - \text{rank}_S(c_i)|
          \end{equation}
          where $\text{rank}_D(c_i)$ and $\text{rank}_S(c_i)$ denote the rank positions
          of category $c_i$ in the overall and filtered distributions, respectively.
          This term amplifies deviation when filtering substantially reorders categories.
    \item $\|\mathbf{v}_D - \mathbf{v}_S\|_2$ quantifies the Euclidean distance between
          normalized distributions.
\end{itemize}
For histogram data (continuous binned distributions), the rank change factor is omitted
($\text{rankSig} = 1$), simplifying to $\text{sig}(S) \cdot \|\mathbf{v}_D - \mathbf{v}_S\|_2$.
High deviation scores indicate that the filter reveals a substantially different pattern from the overall data.

\paragraph{Monotonicity (correlation or temporal trend).}
Monotonicity captures the strength of linear association between two quantitative variables, or temporal trends in time-series data. 
For variables $X$ and $Y$, monotonicity is measured
using the Pearson correlation coefficient:
\begin{equation}
    \text{Monotonicity}(X, Y) = \text{sig} \cdot |\rho(X, Y)|
\end{equation}
where:
\begin{equation}
    \rho(X, Y) = \frac{\text{cov}(X, Y)}{\sigma_X \sigma_Y} = \frac{\mathbb{E}[(X - \mu_X)(Y - \mu_Y)]}{\sigma_X \sigma_Y}
\end{equation}
is the Pearson correlation coefficient, with $\text{cov}(X,Y)$ denoting covariance,
$\mu_X, \mu_Y$ the means, and $\sigma_X, \sigma_Y$ the standard deviations of $X$ and $Y$.
The absolute value $|\rho(X, Y)| \in [0, 1]$ quantifies correlation strength regardless
of direction, with values near 1 indicating strong linear relationships (positive or negative)
and values near 0 indicating weak or no linear association.
\paragraph{Chi-square (categorical association).}
The chi-square test assesses the independence of two categorical variables by comparing observed frequencies in a contingency table to expected frequencies under the null hypothesis of independence.
For two categorical variables $A$ (cardinality $C_A$) and $B$ (cardinality $C_B$)
forming a contingency table with observed counts $O_{ij}$ and expected counts
$E_{ij} = \frac{(R_i)(C_j)}{N}$ (where $R_i$ is the $i$-th row sum, $C_j$ is the
$j$-th column sum, and $N$ is the total count), the chi-square statistic is:
\begin{equation}
    \chi^2 = \sum_{i=1}^{C_A} \sum_{j=1}^{C_B} \frac{(O_{ij} - E_{ij})^2}{E_{ij}}
\end{equation}
Higher $\chi^2$ values indicate stronger deviation from independence, suggesting
significant association between the categorical variables.

\paragraph{Using heterogeneous scores for recommendation.}
The absolute values of these metrics are not directly comparable across all
metric families because they measure different statistical properties. In
\systemname{}, each metric ranks candidates within its applicable
data-pattern/visualization type, and the top candidates from these within-metric
rankings are pooled before semantic and contextual re-ranking.

\section{Prompts for LLM-Agent-Based Knowledge Enhancement}
\label{sec.prompt_llm_knowledge_enhancement}

\lsy{This section documents the prompt template used to enrich database schemas with concise business-oriented descriptions. As described in the main paper, these LLM-generated phrases provide lightweight domain knowledge for both individual columns and broader database topics, which we then encode into embeddings for semantic similarity computation. Making this prompt explicit facilitates reproducibility and clarifies how \systemname{} bridges terse schema headers and higher-level analytical semantics.
}

\begin{casebox}
\textbf{\Large Prompt Template: Knowledge Enhancement}

\vspace{0.3cm}

\textbf{\large AI Data Analyst Context:}
\\
\textbf{Role:} You are a professional data analyst who is an expert in data warehousing and business scenarios. You only provide direct answers as requested.
\vspace{0.3cm}

\textbf{\large For Database Column:}\\
\textbf{Goal:} Your task is to provide a concise, direct business definition phrase for a database column. Do not add any notes, explanations, or prefixes.

\vspace{0.2cm}
Example 1: 

Input: Table = "orders", Column = "order\_date", 

Output: The date an order was placed, used for sales trend analysis.

Example 2: 

Input: Table = "customers", Column = "customer\_name", 

Output: Full name of the customer for identification.

\vspace{0.3cm}

\textbf{\large For Database Topic:}\\
\textbf{Goal:} Your task is to provide a concise, direct business summary phrase for a database or data topic. Do not add any notes, explanations, or prefixes.

\vspace{0.2cm}
Example 1: 

Input: Topic = "customers\_and\_addresses", 

Output: Personal customer information and their associated residential addresses.

Example 2: 

Input: Topic = "flight\_company", 

Output: Operational data for airline companies and their flights.

\end{casebox}

\section{Prompt for Baseline~II (LLM-Direct)}
\label{app:baseline-ii-prompt}

To assess whether the benefits of our structured query recommendation pipeline can be reproduced by direct prompting alone, Baseline~II replaces the entire recommendation engine with a single GPT-4o model. At each turn, the model is provided with the database schema, the user’s full query history and results, and the following prompt to generate next-step query suggestions.

\begin{casebox}
\textbf{\Large Prompt Template: Initial Query Recommendation}

\vspace{0.3cm}

\textbf{\large AI Query Recommendation Context:}
\\
\textbf{Role:} You are a query recommendation specialist. You read database schemas and analyst context, then propose simple, high-value natural language questions for exploratory analysis.
\\
\textbf{Goal:} Analyze the ``\{DB\_ID\}'' database to understand basic information.

\vspace{0.3cm}

\textbf{\large Task Description:}\\
Read the database schema and recommend exactly 5 natural language analysis queries for an analyst who is starting exploration.

\vspace{0.2cm}
\textbf{Database Schema:}

\vspace{0.1cm}
JSON object of the database schema

\vspace{0.3cm}

\textbf{\large Constraints:}
\begin{itemize}
    \item Return exactly 5 recommendations.
    \item Each recommendation must be a single, clear question.
    \item Keep them simple and realistic for text-to-SQL systems.
    \item Do not write SQL.
    \item Cover different useful angles when possible.
\end{itemize}

\vspace{0.3cm}

\textbf{\large Required Output Format:}\\
Please provide your output strictly in the following JSON format:
\begin{verbatim}
{
  "recommendations": [
    "question 1",
    "question 2",
    "question 3",
    "question 4",
    "question 5"
  ]
}
\end{verbatim}
\end{casebox}

\begin{casebox}
\textbf{\Large Prompt Template: Context-Aware Next-step Recommendation}

\vspace{0.3cm}

\textbf{\large AI Query Recommendation Context:}
\\
\textbf{Role:} You are a query recommendation specialist. You read database schemas and analyst context, then propose simple, high-value natural language questions for exploratory analysis.
\\
\textbf{Goal:} Analyze the ``\{DB\_ID\}'' database to understand basic information.

\vspace{0.3cm}

\textbf{\large Task Description:}\\
Read the database schema together with the analyst's current exploration context, and recommend exactly 5 next-step natural language queries.

\vspace{0.2cm}
\textbf{Analyst Memory:}

\vspace{0.1cm}
A list of previously identified insights, e.g.,\\
\textit{
- Customers use different payment methods.\\
- Orders are distributed unevenly across dates.
}

\vspace{0.2cm}
\textbf{Last Query:}

\vspace{0.1cm}
\{last\_query\}

\vspace{0.2cm}
\textbf{Last Query Result Summary:}

\vspace{0.1cm}
\{query\_result\_text\}

\vspace{0.2cm}
\textbf{Database Schema:}

\vspace{0.1cm}
JSON object of the database schema

\vspace{0.3cm}

\textbf{\large Constraints:}
\begin{itemize}
    \item Return exactly 5 recommendations.
    \item Each recommendation must be a single, clear question.
    \item Keep them simple and realistic for text-to-SQL systems.
    \item Do not repeat the last query verbatim.
    \item Do not write SQL.
    \item Prefer useful next steps that build on the analyst's progress.
\end{itemize}

\vspace{0.3cm}

\textbf{\large Required Output Format:}\\
Please provide your output strictly in the following JSON format:
\begin{verbatim}
{
  "recommendations": [
    "question 1",
    "question 2",
    "question 3",
    "question 4",
    "question 5"
  ]
}
\end{verbatim}
\end{casebox}

\section{Prompts for LLM-Agent-Based Simulation and Evaluation}
\label{sec.prompt_llm_eval}

\lsy{To systematically compare \systemname{} against the baselines, we employ an LLM-based agentic framework that simulates analysts interacting with the recommendation system. This section provides the exact prompts used to: (i) decide the agent's first query under a cold-start condition, (ii) choose subsequent actions given evolving context and system recommendations, and (iii) run an independent LLM-as-judge that evaluates the resulting interaction logs along multiple user-experience dimensions. 
}

\begin{casebox}
\textbf{\Large Prompt Template: Cold Start Decision}

\vspace{0.3cm}

\textbf{\large AI Data Analyst Context:}
\\
\textbf{Role:} You are a professional data analyst skilled in exploratory analysis with incomplete information. You can only interact with the system through natural language, including viewing recommended exploration questions.
\\
\textbf{Goal:} Analyze the ``\{DB\_ID\}'' database to understand basic customer information.

\vspace{0.3cm}

\textbf{\large Task Description:}\\
You are connected to the database ``\{DB\_ID\}'' for your first query. The system has provided the database schema and some general initial exploration questions.

\vspace{0.2cm}
\textbf{1. Database Schema (Tables and Columns):}

\vspace{0.1cm}
JSON object of the database schema

\vspace{0.2cm}

\textbf{2. System-Recommended Initial Exploration Questions (Natural Language):}

\vspace{0.1cm}
    List of recommended Natural Language questions, e.g.,\\
    1. Show all customers’ name in the database.\\
    2. How many customers are there? 
    ......

\vspace{0.3cm}

\textbf{\large Decision Options \& Rationale:}\\
Based on the information above, decide on the first natural language query to start your analysis. You have three options:
\begin{itemize}
    \item[A)] \textbf{Choose a Recommendation(CHOOSE RECOMMENDATION):} If one of the recommended questions is a perfect starting point, adopt it directly.
    \item[B)] \textbf{Refine a Recommendation(REFINE RECOMMENDATION):} If a recommended question is in the right direction but not specific enough, you can refine it. For example, by adding grouping, filtering, etc.
    \item[C)] \textbf{Formulate a New Question(FORMULATE NEW):} If none of the recommendations are suitable, or you want to start from a completely different angle, formulate a new question based on the database schema.
\end{itemize}
\vspace{0.3cm}

\textbf{\large Required Output Format:}\\
Please provide your output strictly in the following JSON format:
\begin{verbatim}
{
"evaluation": "Overall assessment of the 
 database schema and initial recommendations.",
"decision_rationale": "Detailed explanation of 
 why you chose A, B, or C.",
"first_action_type": "CHOOSE RECOMMENDATION" | 
                     "REFINE RECOMMENDATION" | 
                     "FORMULATE NEW",
"first_query_text": "This is your first natural 
                     language query."
}
\end{verbatim}
\end{casebox}

\begin{casebox}
\textbf{\Large Prompt Template: Next-step Decision}

\vspace{0.3cm}

\textbf{\large AI Data Analyst Context:}
\\
\textbf{Role:} You are a professional data analyst skilled in exploratory analysis with incomplete information. You can only interact with the system through natural language, including viewing recommended exploration questions.
\\
\textbf{Goal:} Analyze the ``\{DB\_ID\}'' database to understand basic customer information.

\vspace{0.3cm}

\textbf{\large Memory log(What you have learned so far):}\\
Data insights learned in previous rounds.

For example: \textit{"The query result provided basic customer information including names, payment methods, and the dates they became customers. It showed that customers use different payment methods and became customers at various times."}

\vspace{0.3cm}
\textbf{\large Database Schema (For your reference to build queries):}

\vspace{0.1cm}
JSON object of the database schema

\vspace{0.3cm}

\textbf{\large Decision Basis:}

Your last query was: \{last\_query\}.

The system returned the following information:
\begin{itemize}
    \item Query Result: \{query\_result\_text\}
    \item \textbf{Important Constraint:} Focus only on the thematically relevant parts of the query results; ignore irrelevant outliers.
\end{itemize}
Based on your last query, the system recommends the following next steps (Natural Language): \{recs\_text\}

\vspace{0.3cm}
\textbf{Your Task:} Analyze the information above and decide on your next step.

\vspace{0.3cm}
\textbf{\large Decision Options \& Rationale:}
\begin{itemize}
    \item[A)] \textbf{Choose a Recommendation (CHOOSE RECOMMENDATION):} If one of the recommended questions perfectly aligns with your next analysis step, adopt it directly.
    \item[B)] \textbf{Refine a Recommendation (REFINE RECOMMENDATION):} If a recommended question is generally useful, but you want to elaborate on it.
              \textit{For example: The recommendation is ``Show all customers'', but you want to see ``Show customers grouped by city''.}
              In this case, you need to formulate a \textbf{new, more detailed} natural language query and explain why this refinement is necessary.
    \item[C)] \textbf{Formulate a New Question (FORMULATE NEW):} If all recommendations are irrelevant to your line of thought, or you need to explore a completely different direction.
              \textit{For example: You just analyzed the geographical distribution of customers and now want to analyze their other thematic features.}
              In this case, you need to formulate a brand new question and explain why you need to start this new exploration path.
\end{itemize}
\vspace{0.3cm}

\textbf{\large Required Output Format:}\\
Please provide your output strictly in the following JSON format:
\begin{verbatim}
{
"new_insight": "...",
"recommendation_evaluation":"An evaluation of the 
 system's recommendations,explaining if they were 
 useful and why you made the final decision.",
"decision_rationale": "A detailed explanation 
 of why you chose A, B, or C.",
"next_action_type": "CHOOSE RECOMMENDATION" | 
                    "REFINE RECOMMENDATION" | 
                    "FORMULATE NEW",
"next_query_text": "This is your next natural 
                    language query."
}
\end{verbatim}
\end{casebox}

\begin{casebox}
\textbf{\Large Prompt Template: Evaluation}

\vspace{0.3cm}

\textbf{\large AI Evaluator Context:}
\\
\textbf{Role:} You are an expert Human-Computer Interaction (HCI) researcher.
\\
\textbf{Goal:} Your task is to evaluate the performance of a ``Query Recommendation System'' in a simulated exploratory analysis session and fill out a standard user experience questionnaire. Please remain fair and impartial.

\vspace{0.3cm}

\textbf{\large Query Recommendation System Background:}

The system is designed to help users, especially those unfamiliar with the database schema and domain knowledge, perform systematic data exploration by providing step-wise query recommendations.
Its core value lies in guiding users to discover meaningful analytical paths and data insights. You will be provided with the complete interaction log of an AI agent (playing the role of a ``user'') with this system.

\vspace{0.3cm}

\textbf{\large Study Session Information:}

An AI agent acts as a ``user'' to perform an exploratory data analysis task. The agent's characteristics and its complete interaction log with the system are provided below.

\begin{itemize}
    \item AI User Persona: \{agent\_persona\}
    \item AI User Goal: \{agent\_goal\}
    \item Simulation Mode: \{simulation\_mode\}
\end{itemize}

\textbf{Action Type Definitions:}

To help you understand the exploration path, here are the definitions for the ``action\_type'' values:

\begin{itemize}
    \item CHOOSE RECOMMENDATION: The agent directly selects one of the system's recommendations as its next query.
    \item REFINE RECOMMENDATION: The agent modifies or elaborates on one of the system's recommendations to create its next query.
    \item FORMULATE NEW: The agent ignores the system's recommendations entirely and formulates a new query from scratch.
\end{itemize}

\textbf{\large Complete Interaction Log:}

\subsection*{Turn 0}
\hrule 
\vspace{0.5em}

\noindent\textbf{System Recommendations:}
\begin{itemize}
    \item[\textbf{R1:}] {First example recommendation text...}
    \item[\textbf{R2:}] {Second example recommendation text...}
\end{itemize}

\noindent\textbf{Action Type:} {first\_action\_type}
\vspace{0.5em}

\noindent\textbf{Query:}
\begin{quote}
First\_query\_text\_for\_turn\_0...
\end{quote}

\noindent\textbf{Rationale:}
\begin{quote}
Rationale for the first action and query goes here...
\end{quote}

\subsection*{Turn 1}
\hrule
\vspace{0.5em}

\noindent\textbf{Insight from the previous turn:}
\begin{quote}
\textit{This is the new insight discovered from the results of Turn 0.}
\end{quote}

\noindent\textbf{System Recommendations:}
\begin{itemize}
    \item[\textbf{R1:}] {First example recommendation text...}
    \item[\textbf{R2:}] {Second example recommendation text...}
\end{itemize}

\noindent\textbf{Action Type:} {next\_action\_type}
\par\vspace{0.5em}

\noindent\textbf{Query:}
\begin{quote}
Next\_query\_text\_for\_turn\_1...
\end{quote}

\noindent\textbf{Rationale:}
\begin{quote}
Based on the insight from the previous turn, the next step is to...
\end{quote}

\hrule
\vspace{0.3cm}

The Subsequent Turns (until N) ...

\vspace{0.3cm}


\textbf{\large Task:}

Based \textbf{only} on the provided interaction log, please fill out the following questionnaire. For each question (statement), you must provide a score from 1 (Strongly Disagree) to 5 (Strongly Agree), along with a detailed, evidence-based rationale. Your rationale \textbf{must} cite specific events or patterns from the interaction log.

\vspace{0.3cm}

\textbf{\large Questionnaire:}
\begin{itemize}
    \item[\textbf{Q1:}] The system generated suggestions that were topically relevant to the user's domain of interest.
    \item[\textbf{Q2:}] The system provided context-aware suggestions for the user's next exploration step.
    \item[\textbf{Q3:}] The user could easily understand the natural language query suggestions recommended by the system.
    \item[\textbf{Q4:}] The system effectively helped the user decide on the next exploration action.
    \item[\textbf{Q5:}] Through its recommendations, the system helped the user discover previously underexplored data attributes or analytical dimensions.
    \item[\textbf{Q6:}] Guided by the system, the user's exploration path was logically coherent, strategically efficient, and successfully achieved its core analytical goals.
\end{itemize}

\vspace{0.3cm}

\textbf{\large Rating Scale:}
\begin{itemize}
    \item 5 (Strongly Agree): The evidence strongly and consistently supports the statement.
    \item 4 (Agree): The evidence generally supports the statement.
    \item 3 (Neutral): The evidence is mixed, unclear, or insufficient.
    \item 2 (Disagree): The evidence generally contradicts the statement.
    \item 1 (Strongly Disagree): The evidence strongly and consistently contradicts the statement.
\end{itemize}

\textbf{\large Special Instructions:}

\vspace{0.1cm}

For Question Q5: If your score is \textbf{4 or higher}, you must also provide a ``discovered\_insights'' field. This field should be a list of strings, where each string is a specific, interesting insight that was discovered. For example: [``Discovered that most high-value orders are concentrated on weekends'', ``Identified that a specific product's sales have an anomalous peak in a particular season''].

For Question Q6: You must include an ``exploration\_path'' field. This field should be a list of strings, where each string is a concise summary of a key step or sub-topic in the user's analytical journey.
Furthermore, if your rating for Q6 is \textbf{below 4 (Agree)}, please answer the following in a separate ``incoherence\_reason'' field: What specific events or turns in the interaction log caused the exploration path to be less logical or coherent? What was the cause (e.g., sudden change of focus, lack of follow-up, etc.)?

\vspace{0.3cm}

\textbf{\large Required Output Format:}

Please provide your final evaluation in the following JSON format strictly. Ensure you include all questions from Q1 to Q6 and add the special fields according to the instructions above.
\begin{verbatim}
{
"Q1": {{ "score": <1-5>, "rationale": "..." }},
"Q2": {{ "score": <1-5>, "rationale": "..." }},
"Q3": {{ "score": <1-5>, "rationale": "..." }},
"Q4": {{ "score": <1-5>, "rationale": "..." }},
"Q5": {{ "score": <1-5>, "rationale": "...", 
         "discovered_insights": ["...", "..."]}},
"Q6": {{ "score": <1-5>, "rationale": "...", 
         "exploration_path": ["...", "..."], 
         "incoherence_reason": "..." }}
}
\end{verbatim}
\end{casebox}

\section{Prompts for LLM-based Translation between NL and SQL}
\label{sec.prompt_llm_nl2sql}

\lsy{This section presents the prompt templates used for NL2SQL and SQL2NL translation. 
In all few-shot NL2SQL experiments on Spider-dev, the in-context demonstrations are 
randomly sampled from the official Spider training split. For each experiment, we 
uniformly sample $k$ question–SQL pairs ($k=1$ or $k=3$) from the entire training set 
\emph{without replacement} and insert them into the NL2SQL prompt as few-shot examples. 
The selection is purely random and does not use any dev or test data.
 
}

\begin{casebox}
\textbf{\Large Prompt Template: NL2SQL}

\vspace{0.3cm}

\textbf{\large Task:}\\
Given an input question, first create a syntactically correct SQL query to run, then look at the results of the query and return the answer. 
You can order the results by a relevant column to return the most interesting examples in the database.

\textbf{\large Note:}\\
Never query for all the columns from a specific table, only ask for a the few relevant columns.

Pay attention to avoid SQL aliases (e.g., renaming columns,
MIN/MAX/SUM/AVG/COUNT, table names) in SQL Query, e.g., 
\begin{itemize}
    \item[1)] Instead of SELECT SUM(order\_quantity) AS total\_quantity, please use SELECT SUM(order\_quantity) (without "AS" for aliases).
    \item[2)] Instead of SELECT c.customer\_name, ca.address\_type FROM Customers c JOIN Customer\_Addresses ca ON c.customer\_id = ca.customer\_id, \\
    please use SELECT Customers.customer\_name, Customer\_Addresses.address\_type FROM Customers JOIN Customer\_Addresses ON Customers.customer\_id = Customer\_Addresses.customer\_id (without table name aliases)
    \item[C)] Instead of SELECT MAX(active\_from\_date) AS last\_active\_date, \\
    please use SELECT MAX(active\_from\_date) (without "AS" for aliases).
\end{itemize}

Pay attention to use only the column names that you can see in the schema description. 
Be careful to not query for columns that do not exist. Also, pay attention to which column is in which table.\\

\textbf{\large Use the following format:}

Question: Question here.\\
SQLQuery: SQL Query to run.\\
SQLResult: Result of the SQLQuery.\\
Answer: Final answer here.\\

\textbf{\large Few-shot Examples:}\\
Few-shot examples (different schemas) showing the exact format to follow:\\

Example 1 Schema:\\
Table: people (People\_ID, Name, Nationality, Height, ...)\\
Question: Show the number of people in each nationality.\\
SQLQuery: SELECT people.Nationality, COUNT(*) FROM people GROUP BY people.Nationality\\
SQLResult:\\
Answer:\\

Other examples...\\

\textbf{\large Prompt Suffix:}

Only use the following tables:
{table\_info}

Question: {input}

\end{casebox}

\begin{casebox}
\textbf{\Large Prompt Template: SQL2NL}

\vspace{0.3cm}

\textbf{\large Task:}

Please translate the following sql query into natural language to users who do not have sql knowledge and keep it simple. Please only output the natural language result.

\{sql\}

Natural language:\\

\textbf{\large Prompt Suffix:}

\{sql\}

Natural language:
\end{casebox}

\section{Technical Evaluation}
\label{sec.technial_evaluation}


\lsy{We begin by providing a quantitative characterization of the evaluation data. \autoref{tab:spider_dev_complexity} reports the per-query complexity statistics of the Spider-Dev set.

Following this dataset-level analysis, we present the full NL2SQL results under the same experimental setup as in the main paper. We then report the complete set of agent-based evaluation metrics—coverage measures, LLM-as-judge ratings, and hit rates—computed over the eight sampled Spider databases. Finally, we include a categorized NL2SQL error analysis highlighting representative failure patterns of the underlying parser.
}

\begin{table}[htbp]
\setlength{\tabcolsep}{6pt}
\centering
\caption{\lsy{Per-query complexity statistics of the Spider-Dev set. 
The left part reports basic query-level statistics, 
and the right shows the average number of aggregation operations used per query by operator type.}}
\label{tab:spider_dev_complexity}
\small
\begin{tabular}{@{}l c@{\hskip 0.5cm} l c@{}}
\toprule
\textbf{Query Statistics} & \textbf{Value} & \textbf{Aggregation (avg. calls/query)} & \textbf{Value} \\
\midrule
Total number of queries & 1034 & COUNT & 0.4139 \\
Average tables per query & 1.51 & SUM & 0.0338 \\
Average columns per query & 2.73 & AVG & 0.0687 \\
 &  & MIN & 0.0261 \\
 &  & MAX & 0.0406 \\
\bottomrule
\end{tabular}
\end{table}

\lsy{\autoref{tab:spider-selection-appedix} supplements our NL2SQL evaluation 
with the official Spider Exact Match (EM) metric. EM measures structural 
equivalence between generated and gold SQL, offering a stricter criterion 
than execution accuracy (EX). The expanded results show stable EM and EX 
performance across different model backbones. GPT-4o with 1-shot demonstrations
achieves the highest EX, while GPT-4o with 3-shot demonstrations achieves the
highest EM.}

\lsy{\autoref{table:detailed-ratings} provides the complete LLM-as-judge 
scores across all qualitative dimensions. These detailed 
ratings further illustrate the systematic advantages of \systemname{} over the 
two baselines in producing semantically meaningful and context-aware 
recommendations.}

\lsy{\autoref{table:coverage-comparison-full} presents the full set of coverage
metrics used in our agent-based evaluation. These metrics quantify the breadth of
exploration induced by each system—spanning tables, columns, aggregation
operators, and SQL clauses. Overall, \systemname{} exhibits substantially broader
schema coverage and a more diverse use of SQL operators compared to Baseline~I
(Interestingness-only), while Baseline~II (LLM-Direct) provides a stronger
coverage comparator whose differences from \systemname{} are not statistically
significant in the main analysis.}

\lsy{\autoref{table:user-interaction-types} summarizes the agent’s action 
distribution across Choose, Refine, and Formulate decisions. The Hit 
Rate captures how often a recommendation is adopted verbatim. Consistent 
with our qualitative analysis, QRec-NLI yields substantially higher
adoption than Baseline~I and comparable adoption to Baseline~II, indicating
close alignment with the agent's recent analytical focus.}

\lsy{
\autoref{table:nl2sql-error-easy}, \autoref{table:nl2sql-error-medium}, and
\autoref{table:nl2sql-error-hard} summarize the NL2SQL errors generated by
GPT-4o, partitioned by difficulty. Each table presents representative
predicted–gold SQL pairs along with the underlying failure modes. Across
categories, several systematic issues emerge—ranging from projection and
aggregation mistakes to erroneous join-path selection and misaligned
entity grounding—highlighting core limitations of contemporary
LLM-based semantic parsers.}

\begin{table*}[ht]
\centering
\small
\caption{\lsy{Evaluation on Spider-dev under different few-shot settings; for each setting, we run three times and report the mean and standard deviation.}}
\label{tab:spider-selection-appedix}
\setlength{\tabcolsep}{2pt}
\renewcommand{\arraystretch}{1.0}
\begin{tabular}{c|cc|cc|cc}
\toprule
\textbf{Few-shot} & 
\multicolumn{2}{c}{\textbf{GPT-4o}} &
\multicolumn{2}{c}{\textbf{GPT-4o-mini}} & 
\multicolumn{2}{c}{\textbf{GPT-3.5-turbo}} \\
\cmidrule(lr){2-3} \cmidrule(lr){4-5} \cmidrule(lr){6-7}
 & \textbf{EM} & \textbf{EX} & \textbf{EM} & \textbf{EX} & \textbf{EM} & \textbf{EX} \\
\midrule
0-shot & 42.1 ($\pm$0.7) & 75.3 ($\pm$0.2) & 44.6 ($\pm$0.5) & 76.5 ($\pm$0.3) & 34.8 ($\pm$0.3) & 74.1 ($\pm$0.3) \\
1-shot & 46.6 ($\pm$0.3) & 81.2 ($\pm$0.3)& 45.6 ($\pm$0.6) & 77.7 ($\pm$0.4) & 41.7 ($\pm$0.5) & 76.4 ($\pm$0.3) \\
3-shot & 54.6 ($\pm$0.2) & 78.2 ($\pm$0.1)& 46.4 ($\pm$0.9) & 78.3 ($\pm$0.3) & 42.5 ($\pm$0.4) & 78.0 ($\pm$0.2) \\
\bottomrule
\end{tabular}
\end{table*}

\begin{table*}[ht]
    \centering
    \footnotesize
    \caption{\lsyRev{LLM-as-judge evaluation scores across all databases, comparing QRec-NLI, Baseline~I (Interestingness-only), and Baseline~II (LLM-Direct, implemented with GPT-4o) on overall and per-dimension ratings.}}
    \label{table:detailed-ratings}
\resizebox{\linewidth}{!}{
    \begin{tabular}{
        p{3cm}
        l
        *{7}{c}
    }
        \toprule
        \textbf{Database} & \textbf{Model} &
        \textbf{Avg. Score} &
        \textbf{Relevance} &
        \textbf{Context} &
        \textbf{Clarity} &
        \textbf{Next Step} &
        \textbf{Discovery} &
        \textbf{Guidance} \\
        \midrule
        \multirow{3}{*}{customer\_complaints}
            & QRec-NLI & 4.83 & 5.00 & 5.00 & 5.00 & 5.00 & 4.00 & 5.00 \\
            & Baseline~I & 3.00 & 3.00 & 2.00 & 5.00 & 2.00 & 3.00 & 3.00 \\
            & Baseline~II   & 3.67 & 4.00 & 4.00 & 4.00 & 4.00 & 3.00 & 3.00 \\
        \midrule
        \multirow{3}{*}{share\_transactions}
            & QRec-NLI & 4.67 & 5.00 & 5.00 & 5.00 & 5.00 & 4.00 & 4.00 \\
            & Baseline~I & 3.83 & 4.00 & 3.00 & 5.00 & 4.00 & 3.00 & 4.00 \\
            & Baseline~II  & 4.00 & 4.00 & 5.00 & 5.00 & 4.00 & 3.00 & 3.00 \\
        \midrule
        \multirow{3}{*}{movie\_1}
            & QRec-NLI & 4.83 & 5.00 & 5.00 & 5.00 & 5.00 & 4.00 & 5.00 \\
            & Baseline~I & 3.33 & 3.00 & 2.00 & 5.00 & 3.00 & 3.00 & 4.00 \\
            & Baseline~II   & 3.17 & 4.00 & 3.00 & 5.00 & 3.00 & 2.00 & 2.00 \\
        \midrule
        \multirow{3}{*}{software\_problems}
            & QRec-NLI & 5.00 & 5.00 & 5.00 & 5.00 & 5.00 & 5.00 & 5.00 \\
            & Baseline~I & 4.00 & 5.00 & 3.00 & 5.00 & 4.00 & 3.00 & 4.00 \\
            & Baseline~II   & 3.83 & 5.00 & 4.00 & 4.00 & 4.00 & 3.00 & 3.00 \\
        \midrule
        \multirow{3}{*}{candidate\_poll}
            & QRec-NLI & 4.50 & 5.00 & 5.00 & 5.00 & 5.00 & 3.00 & 4.00 \\
            & Baseline~I & 3.00 & 3.00 & 2.00 & 5.00 & 2.00 & 3.00 & 3.00 \\
            & Baseline~II  & 3.67 & 3.00 & 4.00 & 4.00 & 4.00 & 3.00 & 4.00 \\
        \midrule
        \multirow{3}{*}{formula\_1}
            & QRec-NLI & 4.50 & 5.00 & 5.00 & 5.00 & 5.00 & 3.00 & 4.00 \\
            & Baseline~I & 3.83 & 4.00 & 3.00 & 5.00 & 4.00 & 3.00 & 4.00 \\
            & Baseline~II  & 4.00 & 5.00 & 4.00 & 5.00 & 5.00 & 2.00 & 3.00 \\
        \midrule
        \multirow{3}{*}{college\_1}
            & QRec-NLI & 4.83 & 5.00 & 5.00 & 5.00 & 5.00 & 4.00 & 5.00 \\
            & Baseline~I & 2.33 & 1.00 & 1.00 & 5.00 & 1.00 & 1.00 & 5.00 \\
            & Baseline~II   & 3.67 & 4.00 & 4.00 & 4.00 & 3.00 & 3.00 & 4.00 \\
        \midrule
        \multirow{3}{*}{train\_station}
            & QRec-NLI & 5.00 & 5.00 & 5.00 & 5.00 & 5.00 & 5.00 & 5.00 \\
            & Baseline~I & 4.33 & 5.00 & 4.00 & 5.00 & 5.00 & 3.00 & 4.00 \\
            & Baseline~II   & 4.00 & 5.00 & 4.00 & 5.00 & 4.00 & 3.00 & 3.00 \\
        \bottomrule
    \end{tabular}}
\end{table*}

\begin{table*}[ht]
    \centering
    \small
    \caption{\lsyRev{Performance comparison of QRec-NLI and Baselines on various coverage metrics.}}
    \label{table:coverage-comparison-full}
    \setlength{\tabcolsep}{10pt} 
    \begin{tabular}{p{3cm}  l  cccc}
        \toprule
        \textbf{Database} & \textbf{Model} & \textbf{Table} & \textbf{Column} & \textbf{Agg.} & \textbf{Clause} \\
         & & \textbf{Cov.} & \textbf{Cov.} & \textbf{Cov.} & \textbf{Cov.} \\
        \midrule

        \multirow{3}{*}{customer\_complaints} 
            & QRec-NLI  & 1.00 & 0.62 & 0.20 & 0.38 \\
            & Baseline~I & 0.75 & 0.38 & 0.40 & 0.25 \\
            & Baseline~II & 0.33   & 0.36   & 0.40   & 0.38   \\
        \midrule

        \multirow{3}{*}{share\_transactions} 
            & QRec-NLI  & 1.00 & 0.93 & 0.40 & 0.13 \\
            & Baseline~I & 0.57 & 0.47 & 0.00 & 0.13 \\
            & Baseline~II & 1.00   & 0.73   & 0.60   & 0.50   \\
        \midrule

        \multirow{3}{*}{movie\_1} 
            & QRec-NLI  & 1.00 & 0.75 & 0.60 & 0.50 \\
            & Baseline~I & 0.67 & 0.50 & 0.00 & 0.13 \\
            & Baseline~II & 1.00   & 0.64   & 0.40   & 0.50   \\
        \midrule

        \multirow{3}{*}{software\_problems} 
            & QRec-NLI  & 1.00 & 0.75 & 0.00 & 0.13 \\
            & Baseline~I & 0.50 & 0.42 & 0.20 & 0.13 \\
            & Baseline~II & 1.00   & 0.81   & 0.20   & 0.25   \\
        \midrule

        \multirow{3}{*}{candidate\_poll} 
            & QRec-NLI  & 1.00 & 0.92 & 0.60 & 0.38 \\
            & Baseline~I & 1.00 & 0.69 & 0.20 & 0.25 \\
            & Baseline~II & 0.67   & 0.54   & 0.40   & 0.38   \\
        \midrule

        \multirow{3}{*}{formula\_1} 
            & QRec-NLI  & 0.38 & 0.24 & 0.40 & 0.38 \\
            & Baseline~I & 0.46 & 0.18 & 0.20 & 0.25 \\
            & Baseline~II & 0.52   & 0.27   & 0.40   & 0.50   \\
        \midrule

        \multirow{3}{*}{college\_1} 
            & QRec-NLI  & 1.00 & 0.59 & 0.20 & 0.25 \\
            & Baseline~I & 0.29 & 0.09 & 0.00 & 0.00 \\
            & Baseline~II & 0.67   & 0.54   & 0.40   & 0.38   \\
        \midrule

        \multirow{3}{*}{train\_station} 
            & QRec-NLI  & 1.00 & 0.91 & 0.20 & 0.38 \\
            & Baseline~I & 1.00 & 0.91 & 0.00 & 0.13 \\
            & Baseline~II & 1.00   & 1.00   & 0.20   & 0.38   \\

        \bottomrule
    \end{tabular}
\end{table*}

\begin{table*}[t]
\centering
\footnotesize
\caption{\lsyRev{Distribution of interaction types (\textit{Choose}, \textit{Refine}, \textit{Formulate}) for QRec-NLI and Baselines across six-action sessions. The Hit Rate, expressed as a percentage, denotes the proportion of direct \textit{Choose} actions.}}
\label{table:user-interaction-types}
\begin{tabular}{l|l|c|c|c|c}
\toprule
\textbf{Database} & \textbf{Model} & \textbf{Choose} & \textbf{Refine} & \textbf{Formulate} & \textbf{Hit Rate} \\
\midrule

\multirow{3}{*}{customer\_complaints}
& QRec-NLI & 5 & 1 & 0 & 83.33 \\
& Baseline~I  & 3 & 1 & 2 & 50.00 \\
& Baseline~II   & 4 & 1 & 1 & 66.67 \\
\midrule

\multirow{3}{*}{share\_transactions}
& QRec-NLI & 4 & 2 & 0 & 66.67 \\
& Baseline~I  & 1 & 5 & 0 & 16.67 \\
& Baseline~II   & 5 & 1 & 0 & 83.33 \\
\midrule

\multirow{3}{*}{movie\_1}
& QRec-NLI & 6 & 0 & 0 & 100.00 \\
& Baseline~I & 3 & 0 & 3 & 50.00 \\
& Baseline~II  & 5 & 1 & 0 & 83.33 \\
\midrule

\multirow{3}{*}{software\_problems}
& QRec-NLI & 3 & 3 & 0 & 50.00 \\
& Baseline~I  & 3 & 3 & 0 & 50.00 \\
& Baseline~II  & 4 & 1 & 1 & 66.67 \\
\midrule

\multirow{3}{*}{candidate\_poll}
& QRec-NLI & 6 & 0 & 0 & 100.00 \\
& Baseline~I & 2 & 0 & 4 & 33.33 \\
& Baseline~II  & 6 & 0 & 0 & 100.00 \\
\midrule

\multirow{3}{*}{formula\_1}
& QRec-NLI & 6 & 0 & 0 & 100.00 \\
& Baseline~I & 4 & 1 & 1 & 66.67 \\
& Baseline~II & 5 & 1 & 0 & 83.33 \\
\midrule

\multirow{3}{*}{college\_1}
& QRec-NLI & 6 & 0 & 0 & 100.00 \\
& Baseline~I & 0 & 0 & 6 & 0.00 \\
& Baseline~II  & 5 & 1 & 0 & 83.33 \\
\midrule

\multirow{3}{*}{train\_station}
& QRec-NLI & 5 & 1 & 0 & 83.33 \\
& Baseline~I & 4 & 2 & 0 & 66.67 \\
& Baseline~II & 4 & 1 & 1 & 66.67 \\
\bottomrule
\end{tabular}
\end{table*}

\begin{table}[ht]
\centering
\caption{NL2SQL Error Summary — Easy Category}
\label{table:nl2sql-error-easy}
\small
\setlength{\tabcolsep}{3pt}
\begin{tabular}{@{}>{\raggedright\arraybackslash}p{0.18\linewidth} >{\raggedright\arraybackslash}p{0.55\linewidth} >{\raggedright\arraybackslash}p{0.19\linewidth}@{}}
\toprule
\textbf{Error Type} & \textbf{Pred / Gold SQL Examples} & \textbf{Cause} \\
\midrule

Projection mismatch (extra or missing columns)
& \textbf{Pred:} SELECT Title, Original\_air\_date FROM Cartoon WHERE Directed\_by='Ben Jones' \newline
  \textbf{Gold:} SELECT Title FROM Cartoon WHERE Directed\_by="Ben Jones";
& Returned attributes not required by the query semantics. \\
\midrule

Incorrect grouping key
& \textbf{Pred:} SELECT id FROM TV\_Channel GROUP BY id HAVING COUNT(*) > 2 \newline
  \textbf{Gold:} SELECT id FROM tv\_channel GROUP BY country HAVING count(*) > 2
& Grouped by the wrong semantic attribute. \\
\midrule

Count vs list confusion
& \textbf{Pred:} SELECT DISTINCT GovernmentForm FROM country WHERE Region='Africa' \newline
  \textbf{Gold:} SELECT COUNT(DISTINCT GovernmentForm) FROM country WHERE Continent="Africa"
& Returned a list instead of the aggregated count. \\
\midrule

Wrong identifier field (code vs name)
& \textbf{Pred:} SELECT Continent FROM country WHERE Code='AIA' \newline
  \textbf{Gold:} SELECT Continent FROM country WHERE Name="Anguilla"
& Used the wrong entity identifier. \\
\midrule

Distinct-count vs total-count confusion
& \textbf{Pred:} SELECT COUNT(DISTINCT state) FROM AREA\_CODE\_STATE \newline
  \textbf{Gold:} SELECT COUNT(*) FROM area\_code\_state
& Misinterpreted the need for distinct counting. \\
\bottomrule
\end{tabular}
\end{table}

\begin{table}[ht]
\centering
\caption{NL2SQL Error Summary — Medium Category}
\label{table:nl2sql-error-medium}
\small
\setlength{\tabcolsep}{3pt}
\begin{tabular}{@{}>{\raggedright\arraybackslash}p{0.18\linewidth} >{\raggedright\arraybackslash}p{0.55\linewidth} >{\raggedright\arraybackslash}p{0.19\linewidth}@{}}
\toprule
\textbf{Error Type} & \textbf{Pred / Gold SQL Examples} & \textbf{Cause} \\
\midrule

Incorrect join path
& \textbf{Pred:} TV\_series JOIN Cartoon ON Episode = Title \newline
  \textbf{Gold:} TV\_Channel.id = TV\_series.Channel
& Followed an invalid join condition. \\
\midrule

Using MAX instead of ORDER BY + LIMIT 1
& \textbf{Pred:} SELECT MAX(transcript\_date) FROM Transcripts \newline
  \textbf{Gold:} SELECT transcript\_date FROM Transcripts ORDER BY transcript\_date DESC LIMIT 1
& Aggregation lost row-level context. \\
\midrule

Concept confusion (Region vs Continent)
& \textbf{Pred:} WHERE Region='Asia' \newline
  \textbf{Gold:} WHERE Continent="Asia"
& Chose a semantically incorrect attribute. \\
\midrule

Overly broad substring matching
& \textbf{Pred:} GovernmentForm LIKE '\%Republic\%' 
\newline
  \textbf{Gold:} GovernmentForm="Republic"
& Used imprecise pattern matching. \\
\midrule

Incorrect entity alignment (code vs name)
& \textbf{Pred:} WHERE CountryCode='ABW' \newline
  \textbf{Gold:} JOIN country ON Name="Aruba"
& Wrong entity resolution. \\
\midrule

Missing cross-table alignment for filtering
& \textbf{Pred:} WHERE CountryCode='AFG' AND IsOfficial='T' \newline
  \textbf{Gold:} JOIN country ON Name="Afghanistan" AND IsOfficial="T"
& Filtered without cross-table mapping from name. \\
\bottomrule
\end{tabular}
\end{table}

\begin{table}[ht]
\centering
\caption{NL2SQL Error Summary — Hard Category}
\label{table:nl2sql-error-hard}
\small
\setlength{\tabcolsep}{3pt}
\begin{tabular}{@{}>{\raggedright\arraybackslash}p{0.18\linewidth} >{\raggedright\arraybackslash}p{0.55\linewidth} >{\raggedright\arraybackslash}p{0.19\linewidth}@{}}
\toprule
\textbf{Error Type} & \textbf{Pred / Gold SQL Examples} & \textbf{Cause} \\
\midrule

Missing LIMIT for superlatives
& \textbf{Pred:} ORDER BY COUNT(*) DESC \newline
  \textbf{Gold:} ORDER BY COUNT(*) DESC LIMIT 1
& Did not operationalize “most/least” as top-1. \\
\midrule

Incorrect anti-join / set-difference
& \textbf{Pred:} LEFT JOIN ... WHERE Cartoon.Directed\_by <> 'Ben Jones' \newline
  \textbf{Gold:} EXCEPT / NOT EXISTS
& WHERE clause nullifies LEFT JOIN semantics. \\
\midrule

Wrong join key in set difference
& \textbf{Pred:} series\_name NOT IN (SELECT Title ...) \newline
  \textbf{Gold:} JOIN via channel-id foreign key
& Compared unrelated textual attributes. \\
\midrule

Incorrect handling of ties for maxima
& \textbf{Pred:} WHERE Height = (SELECT MAX(Height)...) \newline
  \textbf{Gold:} ORDER BY Height DESC LIMIT 1
& Using equality with MAX returns all ties. \\
\bottomrule
\end{tabular}
\end{table}

\clearpage




\section{Correctness Evaluation of Semantic Descriptions}
\label{appendix.correctness_description}
We evaluate the generated semantic descriptions on the 10 databases introduced above, considering only column-level entries of the form \texttt{table: column}, which yields 332 evaluation items in total. Each item is assessed by two independent human experts recruited from the Department of Computer Science at a local university. For each item, annotators are provided with schema-grounded evidence, including the database identifier, target table and column names, column type, primary-key status and foreign-key relations. They are instructed to rely exclusively on this schema evidence and to treat any unsupported business semantics as factual errors.

We report one metric: Correctness (FC) measures whether a description is semantically consistent with the provided schema evidence. In our setting, hallucination is operationalized as unsupported semantic content. Descriptions that introduce business meaning, functional roles, or other semantics not warranted by the schema evidence are penalized, with clearly unsupported content receiving the lowest score.
\begin{itemize}
        \item \textbf{2}: The description is fully correct and contains no unsupported content.
        \item \textbf{1}: The description is largely correct, but contains minor vagueness or slight extrapolation.
        \item \textbf{0}: The description contradicts the schema evidence or introduces unsupported content.
\end{itemize}

Table~\ref{tab:semantic-correctness-human} summarizes the correctness evaluation of the generated semantic descriptions across the 10 databases. Overall, both human experts assign relatively high FC scores, suggesting that the generated descriptions are generally grounded in the schema evidence and often capture the intended column semantics. Performance is strong on several databases, including \textit{movie\_1}, \textit{train\_station}, \textit{customers\_and\_addresses}, and \textit{college\_1}, while lower scores on \textit{candidate\_poll} indicate that schema-grounded description generation remains challenging for some domains.

The two human experts also show reasonably strong consistency in their judgments. As reported in Table~\ref{tab:semantic-correctness-human}, they achieve 81.9\% exact agreement at the item level, together with a moderate positive rank association (Spearman's $\rho = 0.61$). 

Table~\ref{tab:semantic-error-examples} presents representative low-scoring cases from the human evaluation. These cases suggest two recurring failure modes. First, the model sometimes performs unsupported semantic enrichment, mapping generic attributes to domain-specific roles or usage scenarios, such as recruitment-related qualifications or shipping and billing purposes. Second, it occasionally misinterprets attribute type, for example by recasting a competition number or other literal field as a record identifier.


\begin{table*}[ht]
\centering
\small
\setlength{\tabcolsep}{6pt}
\caption{\lsyRev{Correctness evaluation of semantic descriptions. We report the factual correctness (FC) scores assigned by two independent experts for each database, as well as their item-level agreement on the 332 semantic descriptions. Overall is the unweighted mean across the 10 databases. Exact agreement denotes the proportion of items receiving the same FC score from both evaluators. Spearman's $\rho$ is reported as a complementary measure of score association.}}
\label{tab:semantic-correctness-human}
\begin{tabular}{l|c|c}
\toprule
\multicolumn{3}{c}{\textbf{Per-database correctness evaluation}} \\
\midrule
\textbf{Database} & \textbf{Human-Expert-1} & \textbf{Human-Expert-2} \\
\midrule
customer\_complaints           & 1.60 & 1.70 \\
tracking\_share\_transactions  & 1.60 & 1.95 \\
movie\_1                       & 1.60 & 2.00 \\
tracking\_software\_problems   & 1.75 & 1.93 \\
candidate\_poll                & 1.00 & 1.43 \\
formula\_1                     & 1.79 & 1.90 \\
college\_1                     & 1.81 & 1.93 \\
train\_station                 & 1.71 & 2.00 \\
customers\_and\_addresses      & 1.71 & 1.97 \\
student\_transcripts\_tracking & 1.72 & 1.87 \\
\midrule
\textbf{Overall} & \textbf{1.63} & \textbf{1.87} \\
\bottomrule
\end{tabular}

\vspace{0.6em}

\begin{tabular}{lccc}
\toprule
\multicolumn{3}{c}{\textbf{Inter-annotator agreement on semantic descriptions}} \\
\midrule
\textbf{Evaluator Pair} & \textbf{Exact Agreement} & \textbf{Spearman's $\rho$}&  \\
\midrule
Human-Expert-1 vs Human-Expert-2 & 81.9\% & 0.61  \\
\bottomrule
\end{tabular}
\end{table*}

\begin{table*}[ht]
\centering
\footnotesize
\caption{\lsyRev{Representative low-scoring cases in the human evaluation, illustrating common failure patterns in generated semantic descriptions.}}
\label{tab:semantic-error-examples}
\resizebox{\textwidth}{!}{
\begin{tabular}{p{2.6cm}|p{2.4cm}|p{4.8cm}|p{6.0cm}}
\toprule
Database & Attribute & Generated Description & Main Issue \\
\midrule
candidate\_poll 
& \texttt{candidate: consider rate} 
& Percentage indicating candidate suitability or advancement likelihood in the hiring process. 
& Incorrect domain grounding: the description maps a polling attribute to a recruitment-related notion of candidate suitability. \\
\cmidrule(lr){1-4}

candidate\_poll 
& \texttt{candidate: unsure rate} 
& Proportion of evaluators indicating uncertainty about the candidate's qualifications. 
& Unsupported semantic enrichment: the description introduces qualification-related semantics that are not grounded in the schema. \\
\cmidrule(lr){1-4}

formula\_1 
& \texttt{qualifying: number} 
& Unique identifier assigned to a record undergoing qualification review. 
& Attribute-type misinterpretation: a competition number is incorrectly reformulated as a record identifier. \\
\cmidrule(lr){1-4}

formula\_1 
& \texttt{results: number} 
& Unique identifier assigned to a result record for tracking and reference. 
& Attribute-type misinterpretation: the field is recast as a record identifier rather than a competition number. \\
\cmidrule(lr){1-4}

student\_transcripts
\_tracking 
& \texttt{addresses: line} 
& Street address information for shipping and billing purposes. 
& Partially correct but over-specified: the basic address meaning is plausible, but the stated shipping/billing purpose is not supported by schema evidence. \\
\bottomrule
\end{tabular}
}
\end{table*}

\section{Generation Stability Analysis}
\label{appendix.stability_description}
We assess generation stability by repeating semantic description generation three times on the same ten databases and comparing the resulting column-level descriptions of the form \texttt{table: column}. The analysis covers 332 attributes in total, each with three generated descriptions.

For an attribute $a$, let $\{d_a^{(i)}\}_{i=1}^{K}$ denote its generated descriptions across $K$ runs, and let
$
P_a=\{(i,j)\mid 1\le i<j\le K\}
$
be the set of run pairs. We compute pairwise similarity using three complementary metrics. With sentence encoder $E(\cdot)$, sentence-embedding cosine similarity is
\begin{equation}
s_{\cos}(x,y)=
\frac{E(x)^\top E(y)}{\|E(x)\|\,\|E(y)\|}.
\label{eq:cosine-sim}
\end{equation}

For BERTScore, which measures semantic similarity through contextualized token matching and is therefore relatively robust to paraphrastic variation, let $\{\hat{\mathbf{h}}_u\}_{u=1}^{m}$ and $\{\hat{\mathbf{g}}_v\}_{v=1}^{n}$ be the normalized contextual token embeddings of descriptions $x$ and $y$. Its precision, recall, and $F_1$ are
\begin{align}
P_{\mathrm{B}}(x,y) &= \frac{1}{m}\sum_{u=1}^{m}\max_{1\le v\le n}\hat{\mathbf{h}}_u^\top \hat{\mathbf{g}}_v,
\label{eq:bertscore-precision} \\
R_{\mathrm{B}}(x,y) &= \frac{1}{n}\sum_{v=1}^{n}\max_{1\le u\le m}\hat{\mathbf{h}}_u^\top \hat{\mathbf{g}}_v,
\label{eq:bertscore-recall}
\end{align}
\begin{equation}
s_{\mathrm{B}}(x,y)=F_{\mathrm{B}}(x,y)=
\frac{2P_{\mathrm{B}}(x,y)R_{\mathrm{B}}(x,y)}
{P_{\mathrm{B}}(x,y)+R_{\mathrm{B}}(x,y)}.
\label{eq:bertscore-f1}
\end{equation}

For ROUGE-L, which measures surface-form similarity based on longest common subsequence overlap and is therefore more sensitive to wording and word-order variation, let $\operatorname{LCS}(x,y)$ denote the longest common subsequence length between tokenized descriptions $x$ and $y$. Its precision, recall, and $F_1$ are
\begin{align}
P_{\mathrm{L}}(x,y) &= \frac{\operatorname{LCS}(x,y)}{|x|},
\label{eq:rougel-precision} \\
R_{\mathrm{L}}(x,y) &= \frac{\operatorname{LCS}(x,y)}{|y|},
\label{eq:rougel-recall}
\end{align}
\begin{equation}
s_{\mathrm{L}}(x,y)=F_{\mathrm{L}}(x,y)=
\frac{2P_{\mathrm{L}}(x,y)R_{\mathrm{L}}(x,y)}
{P_{\mathrm{L}}(x,y)+R_{\mathrm{L}}(x,y)},
\label{eq:rougel-f1}
\end{equation}
where $|x|$ and $|y|$ denote the number of tokens in the tokenized descriptions.

We then aggregate each metric in a shared way at the attribute level:
\begin{equation}
S_m(a)=\frac{1}{|P_a|}\sum_{(i,j)\in P_a}s_m\!\left(d_a^{(i)},d_a^{(j)}\right),
\qquad
m\in\{\cos,\mathrm{B},\mathrm{L}\}.
\label{eq:attribute-level-stability}
\end{equation}

For a database $D$ with attribute set $A_D$, the reported database-level summary is
\begin{equation}
\bar{S}_m(D)=\frac{1}{|A_D|}\sum_{a\in A_D}S_m(a),
\qquad
m\in\{\cos,\mathrm{B},\mathrm{L}\}.
\label{eq:database-level-stability}
\end{equation}

Table~\ref{tab:semantic-stability} summarizes the cross-run stability of the generated semantic descriptions over three repeated generations. Overall, the descriptions exhibit strong consistency across runs: cosine similarity and BERTScore are high (0.862 and 0.832, respectively), and ROUGE-L also remains substantial at 0.634. Taken together, these results suggest that repeated generations largely preserve the same underlying semantics, while allowing some variation in wording and surface realization. Although the degree of stability varies somewhat across databases, the overall pattern remains consistently positive, indicating that the semantic description generation process is generally stable under repeated prompting. At the same time, this analysis measures cross-run consistency rather than factual validity: a description may be stable across runs while still being unsupported by the schema. We therefore interpret stability as complementary to the correctness evaluation reported in \autoref{appendix.correctness_description}.

We further analyze representative low-stability cases in Table~\ref{tab:semantic-stability-examples}, ranked by BERTScore-$F_1$, and find that instability is primarily concentrated in weakly specified attributes that admit multiple locally plausible interpretations.
The low-stability cases show that when attribute meanings are only weakly specified by the schema, the generated descriptions can still drift across plausible readings. We therefore interpret semantic enhancement as providing useful schema-grounded hints whose meanings are generally stable across runs, while remaining susceptible to ambiguity in underspecified attributes.

\begin{table*}[t]
\centering
\small
\setlength{\tabcolsep}{8pt}
\caption{\lsyRev{Per-database generation stability across three runs. Cos. Sim. denotes $\bar{S}_{\cos}(D)$, BERTScore denotes $\bar{S}_{\mathrm{B}}(D)$, and ROUGE-L denotes $\bar{S}_{\mathrm{L}}(D)$. Overall is the unweighted mean across the 10 databases.}}
\label{tab:semantic-stability}
\begin{tabular}{l|c|c|c}
\toprule
Database & Cos. Sim. & BERTScore & ROUGE-L \\
\midrule
customer\_complaints           & 0.881 & 0.855 & 0.700 \\
tracking\_share\_transactions  & 0.879 & 0.846 & 0.648 \\
movie\_1                       & 0.858 & 0.823 & 0.579 \\
tracking\_software\_problems   & 0.817 & 0.802 & 0.675 \\
candidate\_poll                & 0.834 & 0.796 & 0.572 \\
formula\_1                     & 0.843 & 0.816 & 0.606 \\
college\_1                     & 0.877 & 0.842 & 0.667 \\
train\_station                 & 0.913 & 0.877 & 0.652 \\
customers\_and\_addresses      & 0.872 & 0.841 & 0.651 \\
student\_transcripts\_tracking & 0.849 & 0.817 & 0.585 \\
\midrule
\textbf{Overall}               & \textbf{0.862} & \textbf{0.832} & \textbf{0.634} \\
\bottomrule
\end{tabular}
\end{table*}

\begin{table*}[ht]
\centering
\small
\caption{\lsyRev{Representative low-stability attributes across three runs, ranked by BERTScore-$F_1$ similarity $S_{\mathrm{B}}(a)$.}}
\label{tab:semantic-stability-examples}
\resizebox{\textwidth}{!}{
\begin{tabular}{p{2.9cm}|p{2.7cm}|p{1.7cm}|p{8.2cm}}
\toprule
Database & Attribute & BERTScore & Representative Variation Across Runs \\
\midrule
formula\_1 
& \texttt{qualifying: number} 
& 0.462 
& Cross-run semantic drift between two incompatible interpretations: the attribute is described as a qualification-record identifier in some runs, but as a competitor racing number in another. \\
\midrule
formula\_1 
& \texttt{qualifying: q} 
& 0.534 
& Unstable semantic grounding across runs, alternating between qualifying position, qualification outcome/status, and starting-grid position. \\
\midrule
tracking\_software
\_problems 
& \texttt{staff: other staff details} 
& 0.577 
& Inconsistent specificity across runs, varying between generic administrative metadata, HR-related information, and open-ended free-text notes. \\
\bottomrule
\end{tabular}
}
\end{table*}


\section{Human--LLM Alignment Study}
\label{appendix.human-llm-align}
To validate the reliability of our \textit{LLM-as-Judge} framework, we further conducted a human--LLM alignment study. Specifically, we invited two student evaluators from a local university to independently assess the query logs generated by QRec-NLI using the same rubric as in the previous \textit{LLM-as-Judge} evaluation. Across the 8 databases, this yields 48 rubric items in total, excluding the derived \textit{Avg.} score. To quantify alignment, we report pairwise \textit{Exact Agreement}, Spearman's $\rho$, and assess overall ordinal reliability across all three raters using Krippendorff's $\alpha_{ord}$.

Table~\ref{tab:human_llm_alignment_detailed} shows that the \textit{LLM-as-Judge} is aligned with human evaluation across all 8 databases. The average scores assigned by the three raters are consistently close, and the ratings on most rubric dimensions are nearly identical in most cases. Minor differences mainly arise in the more subjective dimensions, where human raters occasionally assign slightly lower scores than the LLM judge. Overall, the fine-grained results indicate that the LLM judge captures the same evaluation trends as human raters at both the database and rubric levels.

Table~\ref{tab:human_llm_agreement} quantitatively supports the alignment between our \textit{LLM-as-Judge} and human evaluation. The LLM judge achieves high exact agreement with Human-1 and Human-2, reaching 85.4\% and 87.5\%, respectively, together with strong positive rank correlations (Spearman's $\rho = 0.738$ and $\rho = 0.798$).
In addition, the overall ordinal reliability across all three raters is high (Krippendorff's $\alpha_{ord}=0.826$), further suggesting that the evaluation protocol is stable and that the \textit{LLM-as-Judge} provides a reliable approximation of human assessment in our setting.

\begin{table*}[ht]
    \centering
    \footnotesize
    \caption{\lsyRev{Fine-grained results of the Human--LLM Alignment Study for QRec-NLI. For each database, we report rubric-level scores assigned to the QRec-NLI session log by the \textit{LLM-as-Judge}, \textit{Human-1}, and \textit{Human-2}.}}
    \label{tab:human_llm_alignment_detailed}
    \resizebox{\linewidth}{!}{
    \begin{tabular}{
        p{3cm}|
        l|
        *{7}{c}
    }
        \toprule
        \textbf{Database} & \textbf{Rater} &
        \textbf{Avg.} &
        \textbf{Rel.} &
        \textbf{Ctx.} &
        \textbf{Cla.} &
        \textbf{Next} &
        \textbf{Disc.} &
        \textbf{Guid.} \\
        \midrule
        \multirow{3}{*}{customer\_complaints}
            & LLM-as-Judge  & 4.83 & 5.00 & 5.00 & 5.00 & 5.00 & 4.00 & 5.00 \\
            & Human-1       & 4.83 & 5.00 & 5.00 & 5.00 & 5.00 & 4.00 & 5.00 \\
            & Human-2       & 4.67 & 5.00 & 5.00 & 5.00 & 5.00 & 3.00 & 5.00 \\
        \midrule
        \multirow{3}{*}{share\_transactions}
            & LLM-as-Judge  & 4.67 & 5.00 & 5.00 & 5.00 & 5.00 & 4.00 & 4.00 \\
            & Human-1       & 4.83 & 5.00 & 5.00 & 5.00 & 5.00 & 4.00 & 5.00 \\
            & Human-2       & 4.83 & 5.00 & 5.00 & 5.00 & 5.00 & 4.00 & 5.00 \\
        \midrule
        \multirow{3}{*}{movie\_1}
            & LLM-as-Judge  & 4.83 & 5.00 & 5.00 & 5.00 & 5.00 & 4.00 & 5.00 \\
            & Human-1       & 4.67 & 5.00 & 5.00 & 5.00 & 5.00 & 3.00 & 5.00 \\
            & Human-2       & 4.67 & 5.00 & 5.00 & 5.00 & 5.00 & 3.00 & 5.00 \\
        \midrule
        \multirow{3}{*}{software\_problems}
            & LLM-as-Judge  & 5.00 & 5.00 & 5.00 & 5.00 & 5.00 & 5.00 & 5.00 \\
            & Human-1       & 4.83 & 5.00 & 5.00 & 5.00 & 5.00 & 4.00 & 5.00 \\
            & Human-2       & 4.67 & 5.00 & 5.00 & 5.00 & 5.00 & 3.00 & 5.00 \\
        \midrule
        \multirow{3}{*}{candidate\_poll}
            & LLM-as-Judge  & 4.50 & 5.00 & 5.00 & 5.00 & 5.00 & 3.00 & 4.00 \\
            & Human-1       & 4.83 & 5.00 & 5.00 & 5.00 & 5.00 & 4.00 & 5.00 \\
            & Human-2       & 4.67 & 5.00 & 5.00 & 5.00 & 5.00 & 4.00 & 4.00 \\
        \midrule
        \multirow{3}{*}{formula\_1}
            & LLM-as-Judge  & 4.50 & 5.00 & 5.00 & 5.00 & 5.00 & 3.00 & 4.00 \\
            & Human-1       & 4.67 & 5.00 & 5.00 & 5.00 & 5.00 & 4.00 & 4.00 \\
            & Human-2       & 4.50 & 5.00 & 5.00 & 5.00 & 5.00 & 3.00 & 4.00 \\
        \midrule
        \multirow{3}{*}{college\_1}
            & LLM-as-Judge  & 4.83 & 5.00 & 5.00 & 5.00 & 5.00 & 4.00 & 5.00 \\
            & Human-1       & 4.83 & 5.00 & 5.00 & 5.00 & 5.00 & 4.00 & 5.00 \\
            & Human-2       & 4.83 & 5.00 & 5.00 & 5.00 & 5.00 & 4.00 & 5.00 \\
        \midrule
        \multirow{3}{*}{train\_station}
            & LLM-as-Judge  & 5.00 & 5.00 & 5.00 & 5.00 & 5.00 & 5.00 & 5.00 \\
            & Human-1       & 4.83 & 5.00 & 5.00 & 5.00 & 5.00 & 4.00 & 5.00 \\
            & Human-2       & 4.83 & 5.00 & 5.00 & 5.00 & 5.00 & 4.00 & 5.00 \\
        \bottomrule
    \end{tabular}}
\end{table*}

\begin{table}[t]
    \centering
    \footnotesize
    \caption{\lsyRev{Agreement and similarity statistics in the Human--LLM Alignment Study.}}
    \label{tab:human_llm_agreement}
    
    \begin{tabular}{l|c|c}
        \toprule
        \textbf{Rater Pair} & \textbf{Exact Agreement} & \textbf{Spearman's $\rho$} \\
        \midrule
        LLM-as-Judge vs Human-1 & 85.4\% & 0.738 \\
        LLM-as-Judge vs Human-2 & 87.5\%  & 0.798 \\
        Human-1 vs Human-2      & 91.7\%  & 0.942 \\
        \midrule
        \multicolumn{2}{l}{\textbf{Krippendorff's $\alpha_{ord}$ (All Raters)}} & \multicolumn{1}{l}{\textbf{0.826}} \\
        \bottomrule
    \end{tabular}
\end{table}

\clearpage

\section{\texorpdfstring{\lsyRev{Failure Modes Behind the Low Guidance Scores of LLM-Direct}}{Failure Modes Behind the Low Guidance Scores of LLM-Direct}}

\autoref{tab:q7_failure_cases} presents representative failure cases that help explain why Baseline~II (LLM-Direct) receives relatively low scores on the \textit{Guidance} dimension in the \textit{LLM-as-Judge} evaluation. \mrev{In our rubric, \textit{Guidance} measures whether an exploration path remains locally coherent, avoids repetitive or dead-ended turns, and supports progress toward the user's stated analytical goal.} Although LLM-Direct can often generate locally plausible follow-up questions, the examples below show that such plausibility does not necessarily translate into strong multi-step guidance.

A recurring issue is that LLM-Direct tends to remain \mrev{locally reactive rather than consistently extending the exploration}. In \texttt{formula\_1}, the exploration stays thematically relevant, but repeatedly revisits closely related entities such as circuits and constructors without advancing toward richer analytical dimensions such as comparisons, trends, or relationships. In \texttt{movie\_1}, once the interaction encounters an inconsistency around \textit{The Sound of Music}, subsequent turns are largely hijacked by anomaly checking, causing the exploration to drift away from the original goal of understanding the database more broadly. In \texttt{train\_station}, the path begins reasonably with passenger-related analysis, but later turns are diverted to resolving contradictory station counts instead of expanding into other relevant aspects such as platforms, services, and interchanges.

\mrev{Overall, these cases suggest that the low \textit{Guidance} scores of LLM-Direct are mainly caused by three recurring weaknesses: repeated exploration of semantically adjacent topics without sufficient deepening, susceptibility to being derailed by local inconsistencies, and limited ability to recover after such disruptions. In these examples, QRec-NLI keeps the interaction on topic and extends it into new yet relevant dimensions, resulting in exploration paths that receive stronger ratings for local coherence and next-step guidance.}

\begin{table*}[ht]
\centering
\footnotesize
\caption{\lsyRev{Representative failure cases explaining the relatively low \textit{Guidance} scores of Baseline~II (LLM-Direct) in the LLM-as-Judge evaluation.}}
\label{tab:q7_failure_cases}
\renewcommand{\arraystretch}{1.1}
\setlength{\tabcolsep}{5pt}

\newcolumntype{C}[1]{>{\centering\arraybackslash}p{#1}}

\resizebox{\linewidth}{!}{
\begin{tabular}{C{2.4cm} C{0.9cm} p{4.4cm} p{5.0cm} p{5.8cm}}
\toprule
\textbf{Database} & \textbf{Score} & \centering\textbf{Rationale} & \centering\textbf{Exploration Path} & \centering\textbf{Failure Mode} \tabularnewline
\midrule

\texttt{formula\_1} & 3 &
\mrev{The evidence is mixed regarding the local coherence and next-step usefulness of the exploration path. While the path maintains topical consistency by focusing on Formula 1 entities, it shows substantial repetition and limited analytical deepening. The exploration supports basic database familiarization, but does not progress toward richer comparisons, trends, or relationships.}
&
\begin{enumerate}\itemsep0pt \topsep2pt \partopsep0pt \parsep0pt
    \item Explored the number and geographical distribution of Formula 1 circuits
    \item Explored the number and diversity of constructors
    \item Revisited circuit details for broader geographical understanding
    \item Examined constructor nationalities
    \item Returned again to circuit-related information
    \item Repeated constructor nationality analysis
\end{enumerate}
&
The path remains locally relevant but lacks progressive expansion. Closely related entities are revisited multiple times, with minimal new analytical value added in later turns. Instead of moving from entity enumeration to more meaningful dimensions such as trends, comparisons, or relationships, the exploration stays at a shallow descriptive level. This leads to a path that is coherent at the topical level but weak in next-step guidance.
\\
\midrule

\texttt{movie\_1} & 2 &
\mrev{The evidence suggests limited progress toward the stated exploration goal. The stated goal was to understand the basic information of the \texttt{movie\_1} database, but the exploration becomes dominated by repeated investigation of one inconsistency, preventing broader understanding of the database structure, contents, or patterns.}
&
\begin{enumerate}\itemsep0pt \topsep2pt \partopsep0pt \parsep0pt
    \item Started with a year-based movie query to understand inventory
    \item Attempted to explore director information for a discovered movie
    \item Encountered contradictory information about movie existence
    \item Spent multiple turns investigating the inconsistency
    \item Attempted to verify database contents through listing
\end{enumerate}
&
The exploration is derailed by a local anomaly. Once the inconsistency around \textit{The Sound of Music} appears, later turns are devoted almost entirely to anomaly checking rather than systematic database exploration. This causes a clear drift from the original analytical objective and shows limited ability to recover the broader exploration agenda after a local disruption.
\\
\midrule

\texttt{train\_station} & 3 &
\mrev{The evidence is mixed regarding local coherence and goal progress. The exploration starts reasonably with passenger-related analysis and establishes some baseline understanding, but later becomes disrupted by contradictory information about station counts. As a result, the path does not develop into a broader understanding of the database.}
&
\begin{enumerate}\itemsep0pt \topsep2pt \partopsep0pt \parsep0pt
    \item Established baseline passenger totals for stations
    \item Identified top stations by passenger volume
    \item Examined related passenger metrics
    \item Encountered contradictory information about the number of stations
    \item Attempted to verify the inconsistency in later turns
\end{enumerate}
&
\mrev{The path is interrupted by inconsistency resolution. After a reasonable start, the exploration shifts from understanding station characteristics to troubleshooting contradictory station counts. Rather than extending into other relevant dimensions such as platforms, services, or interchanges, later turns focus on reconciliation of conflicting findings. This weakens both next-step usefulness and exploration continuity.}
\\

\bottomrule
\end{tabular}
}
\end{table*}

\end{document}